\documentclass[aps,pra,twocolumn,notitlepage,superscriptaddress,nofootinbib]{revtex4-2}

\usepackage{amsmath,amssymb,mathtools}
\usepackage{bm}              % bold math
\usepackage{siunitx}         % SI units
\usepackage{graphicx}  
\usepackage{physics}      % figures
\usepackage[caption=false]{subfig}

\usepackage{xcolor}   % colour support
\usepackage{hyperref}

\hypersetup{
    colorlinks=true,
    linkcolor=green!60!black,  % ToC, section refs
    urlcolor=green!50!black,   % URLs
    citecolor=green!50!black   % citations
}
\usepackage{cleveref}
\usepackage{tikz}
\usetikzlibrary{calc}

\definecolor{qoWire}{RGB}{50,50,60}
\definecolor{qoBS}{RGB}{25,95,195}
\definecolor{qoDisp}{RGB}{0,120,160}
\definecolor{qoDispL}{RGB}{205,240,250}
\definecolor{qoSq}{RGB}{20,130,80}
\definecolor{qoRot}{RGB}{115,30,160}
\definecolor{qoRotL}{RGB}{238,215,255}
\definecolor{qoSqL}{RGB}{210,245,225}
\definecolor{qoDet}{RGB}{175,35,35}
\definecolor{qoDetL}{RGB}{255,220,220}
\definecolor{qoFB}{RGB}{180,100,0}
\definecolor{qoFBL}{RGB}{255,235,205}
\definecolor{qoNode}{RGB}{20,20,20}

\definecolor{qoWire}{RGB}{50,50,60}
\definecolor{qoBS}{RGB}{25,95,195}
\definecolor{qoSq}{RGB}{20,130,80}
\definecolor{qoRot}{RGB}{115,30,160}
\definecolor{qoRotL}{RGB}{238,215,255}
\definecolor{qoUG}{RGB}{20,130,80}
\definecolor{qoUGL}{RGB}{210,245,225}
\definecolor{qoSqL}{RGB}{210,245,225}
\definecolor{qoDet}{RGB}{175,35,35}
\definecolor{qoDetL}{RGB}{255,220,220}
\definecolor{qoFB}{RGB}{180,100,0}
\definecolor{qoFBL}{RGB}{255,235,205}
\definecolor{qoNode}{RGB}{20,20,20}

\definecolor{qoWire}{RGB}{50,50,60}
\definecolor{qoDisp}{RGB}{0,120,160}
\definecolor{qoDispL}{RGB}{205,240,250}
\definecolor{qoSq}{RGB}{170,140,0}
\definecolor{qoSqL}{RGB}{255,248,200}
\definecolor{qoRot}{RGB}{115,30,160}
\definecolor{qoRotL}{RGB}{238,215,255}
\definecolor{qoFB}{RGB}{180,100,0}
\definecolor{qoNode}{RGB}{20,20,20}

\providecommand{\ArrowD}[2]{\filldraw[qoWire,line width=0.8pt]({#1-0.0879},{#2})--({#1},{#2-0.2109})--({#1+0.0879},{#2})--cycle;}

\newcommand{\ArrowL}[2]{\filldraw[qoWire,line width=0.8pt]({#1},{#2+0.0879})--({#1-0.2109},{#2})--({#1},{#2-0.0879})--cycle;}
\newcommand{\ArrowR}[2]{\filldraw[qoWire,line width=0.8pt]({#1},{#2+0.0879})--({#1+0.2109},{#2})--({#1},{#2-0.0879})--cycle;}

\tikzset{
  W/.style={draw=qoWire,line width=1.2pt},
  cable/.style={draw=qoFB,line width=1.1pt,dashed,->,>=stealth},
}

\makeatletter

\newif\ifequalcontrib@author

\newcommand{\equalcontrib}{\equalcontrib@authortrue}

\def\equalcontrib@text{%
  These authors contributed equally to this work.%
}

\def\equalcontrib@notes@withaffil#1{%
  \ifequalcontrib@author
    \comma@space
    \frontmatter@footnote{\equalcontrib@text}%
    \@if@empty{#1}{}{%
      \comma@space
      \frontmatter@footnote{#1}%
    }%
  \else
    \@if@empty{#1}{}{%
      \comma@space
      \frontmatter@footnote{#1}%
    }%
  \fi
}

\def\equalcontrib@notes@noaffil#1{%
  \ifequalcontrib@author
    \frontmatter@footnote{\equalcontrib@text}%
    \@if@empty{#1}{}{%
      \comma@space
      \frontmatter@footnote{#1}%
    }%
  \else
    \@if@empty{#1}{}{%
      \frontmatter@footnote{#1}%
    }%
  \fi
}

\def\doauthor#1#2#3{%
  \equalcontrib@authorfalse
  \ignorespaces#1\unskip\@listcomma
  \begingroup
    #3%
  \endgroup
    {\equalcontrib@notes@withaffil{#2}}%
    {\equalcontrib@notes@noaffil{#2}}%
  \space\@listand
}

\makeatother

\begin{document}

\title{Local Gaussian bounds on the non-destructive discrimination of two-mode squeezed states}
\author{Mi-Jung So\equalcontrib}
\affiliation{Department of Physics, Korea University, Seoul 02841, Republic of Korea}

\author{James Moran\equalcontrib}
\email{\mbox{jamesmoran@kias.re.kr}}
\affiliation{Quantum Universe Center, Korea Institute for Advanced Study, Seoul 02455, Republic of Korea}

\author{Youngrong Lim}
% \email{\textcolor{red}{X}}
\affiliation{Department of Physics, Chungbuk National University, Cheongju, Chungbuk 28644, Republic of Korea}

\author{Mahn-Soo Choi}
\email{choims@korea.ac.kr}
\affiliation{Department of Physics, Korea University, Seoul 02841, Republic of Korea}

\author{Hyukjoon Kwon}
\email{hjkwon@kias.re.kr}
\affiliation{Quantum Universe Center, Korea Institute for Advanced Study, Seoul 02455, Republic of Korea}
\affiliation{School of Computational Sciences, Korea Institute for Advanced Study, Seoul 02455, Republic of Korea}

\begin{abstract}
Typical measurement setups in quantum systems are destructive, meaning that states are irretrievably altered after measurement. In this work, we analyse non-destructive discrimination of two two-mode squeezed vacuum states using local Gaussian measurements. We investigate a tradeoff relation between the success probability of discrimination and the fidelity of the resulting state with the initial state, and construct a protocol given by local Gaussian measurements, which is optimal within our numerically explored class. We also extend to the case where we allow for additional pre-shared entanglement, and show that this regime allows us to exceed the standard local Gaussian bound for the fidelity-success probability tradeoff. Our work provides a natural extension of the tradeoff between information gain and disturbance in entangled-state discrimination, previously established for finite-dimensional quantum systems, to infinite-dimensional continuous-variable systems.
\end{abstract}

\maketitle
%\tableofcontents

\section{Introduction}
Quantum state discrimination is fundamental to quantum information and quantum communication \cite{Barnett:09,Bae_2015,9781107002173,Pirandola:20,https://doi.org/10.1049/qtc2.12015}. For a set of non-orthogonal states, no measurement can perfectly distinguish between them, and one will always have some non-zero error in deciding which state was measured. The minimum achievable error allowed by quantum mechanics is known as the Helstrom bound \cite{Helstrom:1969aa}.

The standard state discrimination scenario is destructive, meaning that the quantum state is irretrievable post-measurement. For many tasks of practical interest, such as those involving quantum networks, one may wish to preserve the quantum state after discrimination so that its entanglement can be reused \cite{Kimble:2008aa,Kim:2012aa,Kim:aa,Hirche2023quantumnetwork,PhysRevApplied.14.064074,vcrh-hl73}. This leads to the notion of non-destructive discrimination.

Several works have quantified this task in finite-dimensional systems \cite{Satyajit:2018aa,Jain_2009} where the input state is exactly preserved. One may extend this idea by introducing a figure of merit that encodes a tradeoff between the success probability of the discrimination and the fidelity of the remaining quantum state with the input quantum state. In such cases, fundamental bounds can be derived on the maximum value of this tradeoff in the local operations and classical communication (LOCC) regime in terms of the dimension of the system \cite{Bilash:2024aa,Lim:2025aa}.

In continuous-variable (CV) systems, the set of Gaussian local operations and classical communication (GLOCC) as well as Gaussian states are of both theoretical and experimental importance \cite{https://doi.org/10.1049/qtc2.12015,PhysRevA.66.032316,PhysRevLett.89.137903,PhysRevLett.89.137904}. The symplectic structure of Gaussian states and operations~\cite{RevModPhys.84.621} is well-understood and formulated mathematically, and such operations are practical to realise in laboratories. According to the Lloyd-Braunstein theorem~\cite{PhysRevLett.82.1784} on universal quantum computation over continuous variables, non-Gaussian resources are required for true quantum advantage~\cite{RevModPhys.84.621, PhysRevLett.82.1784, PhysRevLett.109.230503}, and this has been demonstrated explicitly in the context of continuous-variable state discrimination~\cite{kennedy,dolinarRe,PhysRevA.54.2728,PhysRevLett.101.210501,Cook:2007aa,Han_2018,PhysRevA.78.022320,Sidhu2023linearoptics,warke2024photonicquantumreceiverattaining,Moran2026nearoptimalcoherent}. Thus studying GLOCC bounds on quantum information tasks allows one to delineate between classically simulable and quantum regimes. Despite the foundational importance of continuous-variable systems, fundamental bounds for the success probability-fidelity tradeoff from finite-dimensional systems are not useful in the CV setup. This is because the bounds obtained in the literature~\cite{Lim:2025aa} are given in terms of the Hilbert space dimension which, when applied to CV systems, yield trivial results.

In this work we address this problem with a ground-up formulation which circumvents a construction in terms of the state space dimension. By focusing on the non-destructive GLOCC discrimination of two-mode squeezed vacuum (TMSV) states with opposite squeezing directions, we establish an explicit trade-off relation between the success probability and the measurement-induced disturbance. Remarkably, for the one-stage GLOCC ($\textrm{GLOCC}_1$) protocol, when the squeezing is sufficiently large, the total score, defined as the product of the success probability and the fidelity between the target and post-measurement states, saturates at $0.626829$. When another TMSV state is available as pre-shared entanglement, we construct a protocol that exceeds the $\textrm{GLOCC}_1$ bound. We also identify a non-trivial strategy that outperforms the trivial measurement-and-repreparation approach when the squeezing of the pre-shared state is lower than that of the target TMSV states to be distinguished. Our results provide the first infinite-dimensional generalisation of the previously established trade-off relation~\cite{Lim:2025aa} between information gain and disturbance in entangled-state discrimination.

The paper is structured as follows. In Sec.~\ref{sec:prelim} we recapitulate the contents of Gaussian states and operations, and local Gaussian discrimination of two TMSVs. In Sec.~\ref{sec:nondest} we formulate a tradeoff relation between success probability and the fidelity of the output state with the input state, we develop the optical circuits which will implement this protocol, and derive the $\textrm{GLOCC}_1$ bound for this problem. Following this, in Sec.~\ref{sec:preshared} we study the role of pre-shared Gaussian entanglement as a resource in this protocol, we demonstrate that any amount of pre-shared entanglement allows one to exceed the $\textrm{GLOCC}_1$ limit for this protocol. Lastly, in Sec.~\ref{sec:conc} we conclude and leave some open questions for future work.
\section{Preliminaries}\label{sec:prelim}
\subsection{Gaussian states}

We briefly review the phase-space description of multimode Gaussian states and fix the notation used throughout this work.

An $n$-mode bosonic system is described by annihilation and creation operators $\hat{a}_k$ and $\hat{a}_k^\dagger$, satisfying $[\hat{a}_k, \hat{a}_l^\dagger] = \delta_{kl}$. The quadrature operators are defined as
\begin{equation}
    \hat{x}_k = \frac{1}{\sqrt{2}}(\hat{a}_k + \hat{a}_k^\dagger), \quad
    \hat{p}_k = \frac{1}{i\sqrt{2}}(\hat{a}_k - \hat{a}_k^\dagger),
\end{equation}
which obey the canonical commutation relations $[\hat{x}_k, \hat{p}_l] = i \delta_{kl}$.
It is convenient to collect the quadrature operators into a vector
\begin{equation}
    \hat{r} = (\hat{x}_1, \hat{p}_1, \dots, \hat{x}_n, \hat{p}_n)^T,
\end{equation}
which satisfies the canonical commutation relations
\begin{equation}
    [\hat{r}, \hat{r}^T] = i \Omega_n := i \bigoplus_{k=1}^n 
    \begin{pmatrix}
        0 & 1 \\
        -1 & 0
    \end{pmatrix},
\end{equation}
where $\Omega_n$ is the symplectic form. 

A Gaussian state is fully characterized by its displacement vector $\bar{r}$ and covariance matrix $\sigma$, defined as
\begin{align}
    \bar{r}_i &= \langle \hat{r}_i \rangle, \\
    \sigma_{ij} &= \frac{1}{2} \langle \{ \hat{r}_i - \bar{r}_i, \hat{r}_j - \bar{r}_j \} \rangle,
\end{align}
where $\{\cdot,\cdot\}$ denotes the anticommutator.

In the Gaussian quantum circuit setting considered in this work, we focus on unitary operations that preserve the state's Gaussian character. These are represented in phase space by a symplectic matrix $S$ together with a displacement vector $d$, such that
\begin{equation}
    \sigma \rightarrow S \sigma S^T, \quad \bar{r} \rightarrow S \bar{r} + d,
\end{equation}
with $S \Omega_n S^T = \Omega_n$.

General local Gaussian operations can be constructed from a set of elementary Gaussian unitaries that serve as building blocks of local Gaussian circuits. We first consider a single-mode displacement operation with complex amplitude $\alpha$, defined by the unitary operator
\begin{equation}
\hat{D}(\alpha) = \exp\!\left(\alpha \hat{a}^\dagger - \alpha^* \hat{a}\right).
\end{equation}
In the phase-space representation, this operation shifts the first moments while leaving the covariance matrix invariant, corresponding to
\begin{equation}
\bar{r} \rightarrow \bar{r} + d, \quad \sigma \rightarrow \sigma,
\end{equation}
where $d = \left(\sqrt{2}\,\mathrm{Re}(\alpha), \sqrt{2}\,\mathrm{Im}(\alpha)\right)^T$ is the corresponding displacement vector.

Next, a phase rotation by an angle $\phi$ acts on a single mode as
\begin{equation}
\hat{a} \to e^{i\phi}\hat{a},
\end{equation}
which corresponds to the symplectic transformation
\begin{equation}
S_{\mathrm{rot}}(\phi)=
\begin{pmatrix}
\cos\phi & \sin\phi \\
-\sin\phi & \cos\phi
\end{pmatrix}.
\end{equation}

Finally, a single-mode squeezing operation with real parameter $s$ is defined by the unitary operator
\begin{equation}
\hat{S}_\mathrm{sq}(s) = \exp\!\left[\frac{s}{2}\left(\hat{a}^2 - \hat{a}^{\dagger 2}\right)\right].
\end{equation}
We parametrize the squeezing by $\gamma=\tanh s$, with $-1<\gamma<1$. 
In the quadrature representation, this corresponds to the symplectic matrix
\begin{equation}
S_{\mathrm{sq}}(\gamma)=
\begin{pmatrix}
\sqrt{\frac{1-\gamma}{1+\gamma}} & 0\\[4pt]
0 & \sqrt{\frac{1+\gamma}{1-\gamma}}
\end{pmatrix}.
\end{equation}

To complete the set of Gaussian operations considered in this work, we introduce the beamsplitter, which is the fundamental two-mode passive transformation. For two modes $\hat a_1$ and $\hat a_2$, the beamsplitter operator is defined as
\begin{equation}
\hat{S}_\mathrm{bs}(\theta) = \exp\!\left[ \frac{\theta}{2} \left( \hat{a}_{1}^\dagger \hat{a}_{2} - \hat{a}_1 \hat{a}_2^\dagger \right) \right],
\end{equation}
where $\theta$ determines the transmissivity of the transformation. The corresponding symplectic matrix is
\begin{equation}
S_{\mathrm{BS}}(\theta) =
\begin{pmatrix}
\cos \frac{\theta}{2}\, \mathbb{I}_2 & \sin \frac{\theta}{2}\, \mathbb{I}_2 \\
-\sin \frac{\theta}{2}\, \mathbb{I}_2 & \cos \frac{\theta}{2}\, \mathbb{I}_2
\end{pmatrix}.
\end{equation}
Here and throughout, $\mathbb{I}_n$ denotes the $n \times n$ identity matrix.

Together with single-mode displacements, rotations, and squeezing operators, beamsplitters form the set of Gaussian transformations sufficient to construct general multi-mode Gaussian unitaries~\cite{PhysRevA.66.032316,PhysRevLett.89.137904,brask2022gaussianstatesoperations}. In this work we consider the class local Gaussian operations with at most one adaptive stage, $\textrm{GLOCC}_1$. If we denote $\textrm{GLOCC}_k$ to be the set of local Gaussian operations with at most $k$ adaptive stages, and GLOCC the fully adaptive protocol, then we have the following
\begin{equation}
  \textrm{GLOCC}_0 \subseteq  \textrm{GLOCC}_1 \subseteq \textrm{GLOCC}_2 \subseteq\ldots \subseteq \textrm{GLOCC}.
\end{equation}

\subsection{Local Gaussian discrimination of two-mode squeezed vacua}
\label{sub:pre_b}
In continuous-variable quantum information processing, TMSV serve as a fundamental resource for bipartite entanglement~\cite{PhysRevLett.112.070402,doi:10.1126/science.282.5389.706,PhysRevA.87.042330}. TMSV are generated by applying the two-mode squeezing operator
\begin{equation}
\hat{S}(r) = \exp\!\left[ r\left( \hat{a}_{1}^{\dagger} \hat{a}_{2}^{\dagger} - \hat{a}_{1} \hat{a}_{2} \right) \right],
\end{equation}
where $r \in \mathbb{R}$ denotes the squeezing parameter. In the phase-space representation, the vacuum corresponds to vanishing displacement and covariance matrix $\sigma = \frac{1}{2}\mathbb{I}_4$.
It is convenient to parametrize TMSV in terms of $\lambda = \tanh r$, with $-1 < \lambda < 1$. Using the identities $\cosh(2r) = \frac{1+\lambda^2}{1-\lambda^2}$ and $\sinh(2r) = \frac{2\lambda}{1-\lambda^2}$, the covariance matrix takes the block form
\begin{equation}
\sigma_{\lambda} =
\frac{1}{2(1-\lambda^2)}
\begin{pmatrix}
(1+\lambda^2)\, \mathbb{I}_2 & 2\lambda\, \mathbb{Z} \\
2\lambda\, \mathbb{Z} & (1+\lambda^2)\, \mathbb{I}_2
\end{pmatrix},
\end{equation}
where $\mathbb{Z} = \mathrm{diag}(1,-1)$.

The problem of discriminating two Gaussian states has been considered by several authors using a variety of distinguishability metrics including the Helstrom bound and the quantum Chernoff bound~\cite{PhysRevA.86.022306,PhysRevA.78.012331}. Here the task is to discriminate the two TMSVs $\ket{+\lambda}$ and $\ket{-\lambda}$, prepared with equal prior probabilities in the single-shot Helstrom scenario via local homodyne detection. 
Let $\mathbf{z} = (x_a,x_b)^T$ collect the measurement outcomes obtained on modes $a$ and $b$. For Gaussian states, the corresponding probability densities take the form
\begin{equation}
p_{\pm}(\mathbf{z}) =
\frac{1}{2\pi \sqrt{\det V_{\pm}}}
\exp\!\left(-\frac{1}{2}\mathbf{z}^T V_{\pm}^{-1} \mathbf{z} \right),
\end{equation}
where $V_{\pm}$ are the covariance matrices associated with the measured quadratures, obtained by restricting the covariance matrix of each candidate state to the measured modes.

The optimal decision rule assigns each outcome to the more likely state, partitioning the outcome space $\mathbb{R}^2$ into the decision regions $\mathbb{D}_{+}=\left\{\mathbf{z}\in \mathbb{R}^2: p_+(\mathbf{z})\geq p_- (\mathbf{z})\right\}$ and $\mathbb{D}_{-}=\left\{\mathbf{z}\in \mathbb{R}^2: p_+(\mathbf{z})< p_- (\mathbf{z})\right\}$.
Averaging over the two equally likely inputs, the success probability is then
\begin{equation}\label{eq:homodyne}
\begin{split}
    P &=\frac{1}{2}\sum_{s=\pm}\int_{\mathbb{D}_s}\mathrm{d}\mathbf{z}\; p_s(\mathbf{z})
    \\&=
\frac{1}{2}
+\frac{1}{4}
\int_{\mathbb{R}^2} \mathrm{d}\mathbf{z}\,
\left| p_{+}(\mathbf{z}) - p_{-}(\mathbf{z}) \right|.
\end{split}
\end{equation}

\section{Local non-destructive Gaussian discrimination of two-mode squeezed states}
\label{sec:nondest}
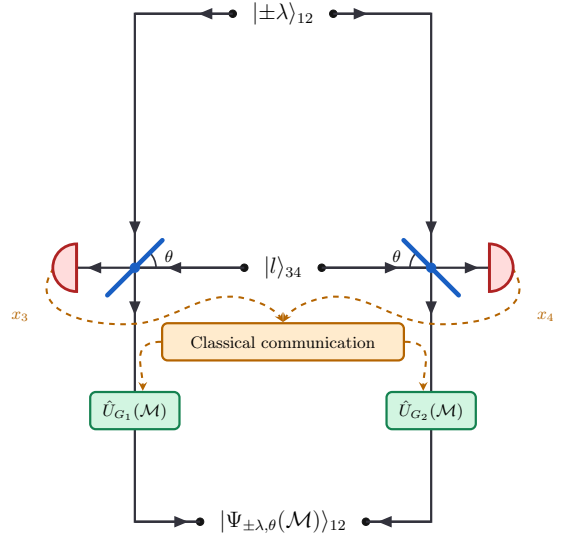
\begin{figure}
 \begin{center}\scalebox{0.7}{
\begin{tikzpicture}[x=1cm,y=1cm]

%% Layout (taller to make room for the CC box row):
%%   yTop=4.8, yBot=-4.8, BS at y=0
%%   CC box at y=-1.4
%%   displacement at y=-2.5, squeezer at y=-3.7

%% ── Top rail ────────────────────────────────────────────────
\draw[W](-2.8,4.8)--(-0.95,4.8); \draw[W](0.95,4.8)--(2.8,4.8);
\filldraw[qoNode](-0.95,4.8)circle(2.0pt); \filldraw[qoNode](0.95,4.8)circle(2.0pt);
\node[font=\large]at(0,4.8){$|{\pm\lambda}\rangle_{12}$};

%% ── Bottom rail ─────────────────────────────────────────────
\draw[W](-2.8,-4.8)--(-1.567,-4.8); \draw[W](1.567,-4.8)--(2.8,-4.8);
\filldraw[qoNode](-1.567,-4.8)circle(2.0pt); \filldraw[qoNode](1.567,-4.8)circle(2.0pt);
\node[font=\large]at(0,-4.8){$|\Psi_{\pm\lambda,\theta}(\mathcal{M})\rangle_{12}$};

%% ── Horizontal rail ─────────────────────────────────────────
\draw[W](-2.8,0)--(-0.7316,0); \draw[W](0.7316,0)--(2.8,0);
\filldraw[qoNode](-0.7316,0)circle(2.0pt); \filldraw[qoNode](0.7316,0)circle(2.0pt);
\node[font=\large]at(0,0){$|l\rangle_{34}$};
\draw[W](-2.8,0)--(-3.88,0); \draw[W](2.8,0)--(3.88,0);

%% ── Vertical arms ───────────────────────────────────────────
\draw[W](-0.95,4.8)--(-2.8,4.8)--(-2.8,-4.8)--(-1.567,-4.8);
\draw[W](0.95,4.8)--(2.8,4.8)--(2.8,-4.8)--(1.567,-4.8);
\draw[W](-2.8,0)--(-0.7316,0); \draw[W](0.7316,0)--(2.8,0);

%% ── Gates: displacement at y=-2.5, squeezer at y=-3.7 ──────

\node[draw=qoUG,fill=qoUGL,line width=1.2pt,rounded corners=3pt,
      minimum width=48pt,minimum height=20pt,font=\small,inner sep=3pt]
      (ugL)at(-2.8,-2.7){$\hat{U}_{G_1}(\mathcal{M})$};
\node[draw=qoUG,fill=qoUGL,line width=1.2pt,rounded corners=3pt,
      minimum width=48pt,minimum height=20pt,font=\small,inner sep=3pt]
      (ugR)at(2.8,-2.7){$\hat{U}_{G_2}(\mathcal{M})$};

%% ── Central classical communication box, own row at y=-1.4 ──
\node[draw=qoFB,fill=qoFBL,line width=1.2pt,rounded corners=3pt,
      minimum width=130pt,minimum height=20pt,font=\small]
      (cc)at(0,-1.4){Classical communication};

%% ── BS arrowheads ───────────────────────────────────────────
\ArrowD{-2.8}{0.8609} \ArrowD{2.8}{0.8609}
\ArrowD{-2.8}{-0.65}  \ArrowD{2.8}{-0.65}
\ArrowL{-1.9391}{0}   \ArrowR{1.9391}{0}
\ArrowL{-3.45}{0}     \ArrowR{3.45}{0}
\ArrowL{-1.35}{4.8}   \ArrowR{1.35}{4.8}
\ArrowR{-2.04}{-4.8}  \ArrowL{2.04}{-4.8}

%% ── θ arcs ──────────────────────────────────────────────────
\draw[qoWire,line width=0.9pt](-2.4,0)arc[start angle=0,end angle=45,radius=0.40];
\node[font=\small]at(-2.15,0.22){$\theta$};
\draw[qoWire,line width=0.9pt](2.4,0)arc[start angle=180,end angle=135,radius=0.40];
\node[font=\small]at(2.15,0.22){$\theta$};

%% ── Beam splitters ──────────────────────────────────────────
\draw[qoBS,line width=2.5pt,line cap=round](-3.32,-0.52)--(-2.28,0.52);
\filldraw[qoBS](-2.8,0)circle(2.5pt);
\draw[qoBS,line width=2.5pt,line cap=round](2.28,0.52)--(3.32,-0.52);
\filldraw[qoBS](2.8,0)circle(2.5pt);

%% ── Feedback: detectors → classical communication box (top centre) ──
%% Curve well below/around the BS (y=0) before heading to (cc.north)
\draw[cable]
  (-4.33,0)
  ..controls(-4.9,-0.85)and(-3.6,-1.15)..
  (-2.7,-1.05)
  ..controls(-2.0,-0.95)and(-1.6,-0.5)..
  (-0.9,-0.55)
  ..controls(-0.4,-0.6)and(-0.1,-0.85)..
  (0,-1.027);
\node[font=\small,color=qoFB,left=1pt]at(-4.65,-0.92){$x_3$};
\draw[cable]
  (4.33,0)
  ..controls(4.9,-0.85)and(3.6,-1.15)..
  (2.7,-1.05)
  ..controls(2.0,-0.95)and(1.6,-0.5)..
  (0.9,-0.55)
  ..controls(0.4,-0.6)and(0.1,-0.85)..
  (0,-1.027);
\node[font=\small,color=qoFB,right=1pt]at(4.65,-0.92){$x_4$};

%% ── Cables from CC box sides into tops of UG gates ──────────
\draw[cable]
  (-2.3058,-1.4)
  ..controls(-2.9,-1.4)and(-2.65,-1.9)..
  (-2.65,-2.3274);
\draw[cable]
  ( 2.3058,-1.4)
  ..controls( 2.9,-1.4)and( 2.65,-1.9)..
  ( 2.65,-2.3274);

%% ── Detectors ───────────────────────────────────────────────
\draw[qoDet,line width=1.5pt,fill=qoDetL](-3.90,0.45)arc[start angle=90,end angle=270,radius=0.45]--cycle;
\draw[qoDet,line width=1.5pt](-3.90,0.45)--(-3.90,-0.45);
\draw[qoDet,line width=1.5pt,fill=qoDetL](3.90,-0.45)arc[start angle=-90,end angle=90,radius=0.45]--cycle;
\draw[qoDet,line width=1.5pt](3.90,0.45)--(3.90,-0.45);

\end{tikzpicture}}
\end{center}
\caption{\label{general_gaussian_circuit}General local Gaussian circuit with one adaptive stage for the non-destructive discrimination of two-mode squeezed vacua. Each party mixes the received signal mode ($1$ or $2$) with an ancillary mode ($3$ or $4$) on a local beamsplitter of angle $\theta$, performs homodyne detection on the ancillary output to obtain $\mathcal{M}=\{x_3,x_4\}$, and—conditioned on the classically communicated outcomes—applies a single-mode Gaussian operation $\hat{U}_{G_i}(\mathcal{M})$ to the surviving signal mode.}
\end{figure}

We consider the task of non-destructive local discrimination of bipartite continuous-variable entangled states shared between two distant parties, Alice and Bob. The goal is to identify the given state using only local operations and classical communication, while preserving the state after the discrimination process. In the scenario considered here, the states to be discriminated are two-mode squeezed vacua (TMSV) with squeezing parameters $\pm \lambda$, denoted by $|+\lambda\rangle$ and $|-\lambda\rangle$. 
A referee, Charlie, chooses one of these states with equal probability and distributes the modes to Alice and Bob. 
The most general local Gaussian strategy we consider is the circuit depicted in Fig.~\ref{general_gaussian_circuit}. 

Each party mixes the received signal mode with an ancillary mode on a local beamsplitter and performs homodyne detection on the ancillary output. The outcomes are collected into the set $\mathcal{M}=\{x_3,x_4\}$ and exchanged by classical communication; conditioned on $\mathcal{M}$, each party applies a local Gaussian operation $\hat{U}_{G_i}(\mathcal{M})$ to its surviving signal mode. The ancillary modes $3$ and $4$ are prepared in a TMSV $\ket{l}_{34}$, whose squeezing parameter $l\in(-1,1)$ quantifies the entanglement (pre-)shared by Alice and Bob before the protocol begins. Throughout Sec.~\ref{sec:nondest} we set $l=0$, so that the ancillae reduce to the vacuum; in Sec.~\ref{sec:preshared} we instead take $l\neq 0$, supplying pre-shared entanglement as a resource, and examine how it improves the discrimination performance. 

The discrimination is now performed on the ancillary modes rather than on the signal itself: the sign of the squeezing is inferred from the pair of homodyne outcomes $\mathbf{z}=\left(x_3, x_4\right)^T$, to which the decision rule of Sec.~\ref{sub:pre_b} applies with the densities $p_\pm$ of the measured ancillary quadratures. 

Conditioned on $\mathcal{M}$, the signal modes $1$ and $2$ are left in the pure state $\ket{\psi_\pm(\mathbf{z})}$ for the input $\ket{\pm \lambda}$, obtained with probability density $p_\pm(\mathbf{z})$, whose fidelity with the input is $F(\lambda, \theta; \mathcal{M})=|\langle \pm \lambda|\psi_\pm (\mathbf{z})\rangle |^2 $. Here $\mathcal{M}$ and $\mathbf{z}$ carry the same data, the former as the classical record conditioning the gates and the latter as the variable of the outcome densities. 

The figure of merit of the protocol is the score function, the average of the product of success and fidelity over a single run,
\begin{equation}\label{eq:score}
    \overline{P\cdot F}=\frac{1}{2}\sum_\pm \int_{\mathbb{D}_\pm}\mathrm{d}\mathbf{z}\; p_\pm (\mathbf{z})\left| \langle \pm \lambda |\psi_\pm (\mathbf{z})\rangle\right|^2, 
\end{equation}
which is the success probability of Eq.~\eqref{eq:homodyne} with each correct identification weighted by the fidelity of the state it leaves behind. Whenever the conditional fidelity is independent of the outcomes and of the sign, $F(\lambda, \theta; \mathcal{M})=F(\lambda,\theta)$, the average factorizes as $\overline{P\cdot F}=P\cdot F$. 

\subsection{Tradeoff between success probability and fidelity}
\begin{figure}
 \begin{center}
 \scalebox{0.7}{
\begin{tikzpicture}[x=1cm,y=1cm]

\tikzset{ccline/.style={draw=qoFB,line width=1.1pt,
  dash pattern=on 4pt off 3pt,dash phase=0.21pt}}
\begin{scope}[yshift=0cm]
  \draw[W](0,1.1)--(0,-1.1);
  \draw[ccline](0.5,0.923)--(0.5,0.76);
  \draw[draw=qoFB,line width=1.1pt,->,>=stealth](0.5,0.76)--(0.5,0.5491);
  \draw[ccline](0.5,0.5491)--(0.5,0.3726);
  \node[draw=qoUG,fill=qoUGL,line width=1.2pt,rounded corners=3pt,
        minimum width=48pt,minimum height=20pt,font=\small,inner sep=3pt]
        at(0,0){$\hat{U}_{G_i}(\mathcal{M})$};
  \ArrowD{0}{0.76}
  \ArrowD{0}{-0.76}
  \foreach \y in {1.25,1.38,1.51}{\filldraw[qoWire](0.25,\y)circle(0.8pt);}
  \foreach \y in {-1.25,-1.38,-1.51}{\filldraw[qoWire](0.25,\y)circle(0.8pt);}
\end{scope}
\node[font=\Large] at(1.7,0){$=$};
\begin{scope}[xshift=3.4cm]
  \draw[W](0,1.1)--(0,-1.1);
  \ArrowD{0}{0.76}
  \ArrowD{0}{-0.76}
  \foreach \y in {1.25,1.38,1.51}{\filldraw[qoWire](0.25,\y)circle(0.8pt);}
  \foreach \y in {-1.25,-1.38,-1.51}{\filldraw[qoWire](0.25,\y)circle(0.8pt);}
\end{scope}

\end{tikzpicture}}
\end{center}
\caption{\label{basic_circuit}Simplest scheme with an initial TMSV state $|\pm\lambda\rangle$ prepared by Charlie and distributed to Alice (left) and Bob (right). Each party mixes the received mode with an auxiliary vacuum mode $|0\rangle$ using a local beamsplitter with parameter $\theta$. 
Both parties then perform homodyne detection on one of the output modes, producing outcomes $x_3$ and $x_4$, while the remaining mode forms the post-measurement state used for the fidelity readout. We collect these outcomes into the classical set $\mathcal{M}=\{x_3, x_4\}$. This corresponds to choosing $l=0$ and $\hat{U}_{G_i}(\mathcal{M})=\mathbb{I}$, the identity operator, in the general circuit.}
\end{figure}
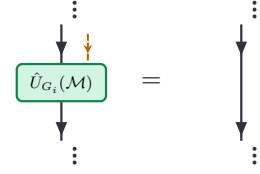

We now build up the optimal local Gaussian strategy in stages, activating the conditional gates $\hat{U}_{G_i}(\mathcal{M})$ of Fig.~\ref{general_gaussian_circuit} one at a time. We begin with the simplest instance, illustrated in Fig.~\ref{basic_circuit}, obtained by setting the ancillary state to the vacuum ($l=0$) and switching the gates off entirely $\hat{U}_{G_i}(\mathcal{M})=\mathbb{I}$, so that only the local beamsplitters and homodyne detection remain. The conditional displacements are switched on later in this subsection, and the conditional squeezing in Sec.~\ref{sub:local_squeezing}. With the gates switched off, the outcomes $\mathcal{M}$ enter only through the decision rule, and the signal modes emerge from the beamsplitters unprocessed. We restrict the local measurements to homodyne detection throughout: although heterodyne and more broadly general-dyne measurements can yield a higher success probability in some parameter regimes and configurations~\cite{Muller:2012aa}, homodyne detection is optimal once the fidelity of the post-measurement state is also taken into account, as shown in Appendix~\ref{app: comparison_of_homodyne_and_heterodyne}. We remark that recent results have found that homodyne detection is the optimal Gaussian measurement to quantify the distinguishability between two Gaussian states in terms of the maximum relative entropy~\cite{turner2026optimaldiscriminationgaussianstates}. For other distinguishability metrics, such as the Helstrom bound, this is in general not true.

\begin{figure}\label{basic_success}
\center
\includegraphics[scale=0.9]{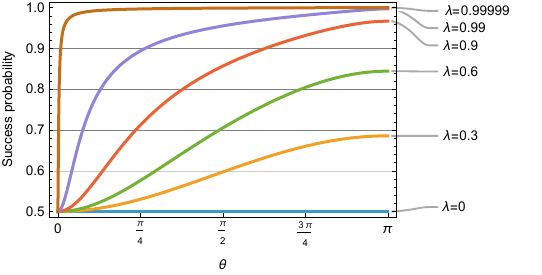}
\caption{Success probabilities of the simplest scheme (Fig.~\ref{basic_circuit}) and the measurement-based displacement circuit (Fig. ~\ref{optimal_displacement_circuit}), as functions of $\theta$ for several values of $\lambda$, which coincide for the same setting with initial state ($|\lambda\rangle$) and two vacuum nodes.}
\end{figure}

The success probability of the simplest scheme shown in Fig.~\ref{basic_circuit}, depends on both the initial state parameter $\lambda$ and the beamsplitter angle $\theta$. We denote it by 
\begin{equation}\label{eq:basic_success}
    P=\frac{1}{2}+\frac{1}{\pi}\arcsin \left[\frac{ \lambda (1-\cos \theta)}{1-\lambda^2 \cos \theta}\right],
\end{equation} 
which makes the explicit dependence on the input state $|\lambda\rangle$ and the beamsplitter angle $\theta$ clear.  Fig.~\ref{basic_success} shows $P$ as a function of $\theta$ for several values of $\lambda$.  On the physical domain $0<\lambda<1$ and $0\leq \theta \leq \pi$, $P$ increases monotonically with each of $\lambda$ and $\theta$, as is apparent from these curves and verified analytically in Appendix~\ref{app:succ}, and it takes the fixed boundary values $P|_{\theta=0}=P|_{\lambda=0}=\frac{1}{2}$ and $P|_{\lambda=1}=1$. 

We next consider the fidelity of the simplest scheme. For a given set of measurement outcomes $\mathcal{M}=\{x_3,x_4\}$, we define the fidelity between the input state and the post-measurement state as $F(\lambda,\theta; \mathcal{M})$. The overall performance of the scheme is characterized by the average fidelity obtained by integrating over all possible outcomes, 
\begin{equation}\label{eq:simple_fid_ave}
    F_{\mathrm{avg}}(\lambda,\theta) = \int \mathrm{d}x_3  \mathrm{d}x_4 \; p(x_3,x_4|\lambda,\theta)  F(\lambda,\theta; x_3, x_4),  
\end{equation} 
where $p(x_3,x_4|\lambda,\theta)$ is the joint probability distribution of the measurement outcomes in the simplest scheme. This integral can be evaluated analytically, yielding 
\begin{equation} \label{eq:simple_fid_closed}
    F_{\mathrm{avg}}(\lambda,\theta) = \frac{1-\lambda^{2}}{1-\tfrac{3}{8}\lambda^{2}-\tfrac{1}{2}\lambda^{2}\cos\theta-\tfrac{1}{8}\lambda^{2}\cos 2\theta}. 
\end{equation}
The average fidelity $F_{\mathrm{avg}}(\lambda,\theta)$ is plotted as a function of $\theta$ for several values of $\lambda$ in Fig.~\ref{optimal_displacement_fidelity}.

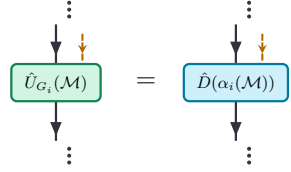
\begin{figure}
 \begin{center}
 \scalebox{0.7}{
\begin{tikzpicture}[x=1cm,y=1cm]

\tikzset{ccline/.style={draw=qoFB,line width=1.1pt,
  dash pattern=on 4pt off 3pt,dash phase=0.21pt}}
%% LHS
\begin{scope}[yshift=0cm]
  \draw[W](0,1.1)--(0,-1.1);
  \draw[ccline](0.5,0.923)--(0.5,0.76);
  \draw[draw=qoFB,line width=1.1pt,->,>=stealth](0.5,0.76)--(0.5,0.5491);
  \draw[ccline](0.5,0.5491)--(0.5,0.3726);
  \node[draw=qoUG,fill=qoUGL,line width=1.2pt,rounded corners=3pt,
        minimum width=48pt,minimum height=20pt,font=\small,inner sep=3pt]
        at(0,0){$\hat{U}_{G_i}(\mathcal{M})$};
  \ArrowD{0}{0.76}
  \ArrowD{0}{-0.76}
  \foreach \y in {1.25,1.38,1.51}{\filldraw[qoWire](0.25,\y)circle(0.8pt);}
  \foreach \y in {-1.25,-1.38,-1.51}{\filldraw[qoWire](0.25,\y)circle(0.8pt);}
\end{scope}
%% "="
\node[font=\Large] at(1.7,0){$=$};
%% RHS: D only
\begin{scope}[xshift=3.4cm]
  \draw[W](0,1.1)--(0,-1.1);
  \draw[ccline](0.5,0.923)--(0.5,0.76);
  \draw[draw=qoFB,line width=1.1pt,->,>=stealth](0.5,0.76)--(0.5,0.5491);
  \draw[ccline](0.5,0.5491)--(0.5,0.3726);
  \node[draw=qoDisp,fill=qoDispL,line width=1.2pt,rounded corners=3pt,
        minimum width=56pt,minimum height=20pt,font=\small,inner sep=3pt]
        at(0,0){$\hat{D}(\alpha_i(\mathcal{M}))$};
  \ArrowD{0}{0.76}
  \ArrowD{0}{-0.76}
  \foreach \y in {1.25,1.38,1.51}{\filldraw[qoWire](0.25,\y)circle(0.8pt);}
  \foreach \y in {-1.25,-1.38,-1.51}{\filldraw[qoWire](0.25,\y)circle(0.8pt);}
\end{scope}
\end{tikzpicture}}
\end{center}
\caption{\label{optimal_displacement_circuit}Measurement-based displacement scheme. Conditioned on the set of measurement outcomes $\mathcal{M}=\{x_3, x_4\}$ obtained on modes $3$ and $4$, the displacement gates $\hat{D}\left(\alpha_i(\mathcal{M})\right)\quad(i=1,2)$ shift the post-measurement state to the phase-space origin, rendering the output fidelity independent of $\mathcal{M}$ and optimal. This corresponds to choosing $l=0$ and $\hat{U}_{G_i}(\mathcal{M})=\hat{D}(\alpha_i (\mathcal{M}))$ in the general circuit.}
\end{figure}

\begin{figure}\label{optimal_displacement_fidelity}
\center
\includegraphics[scale=0.9]{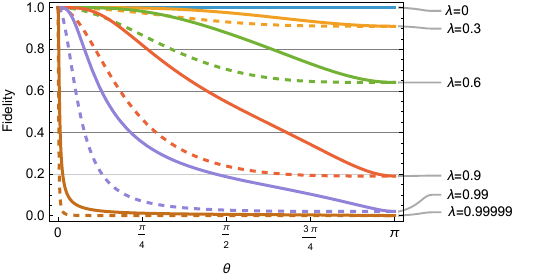}
\caption{Fidelity as a function of $\theta$ for the simplest scheme (Fig.~\ref{basic_circuit}) and the measurement-based displacement circuit (Fig.~\ref{optimal_displacement_circuit}), shown for several values of $\lambda$. Each value of $\lambda$ is represented by a distinct color, with the dashed and solid curves of the same color corresponding to the simplest scheme and the measurement-based displacement circuit, respectively. The latter consistently achieves higher fidelity, demonstrating the advantage of this approach.}
\end{figure}

For the simplest scheme, the fidelity $F(\lambda,\theta; \mathcal{M})$ depends on the measurement outcomes $\mathcal{M}=\{x_3, x_4\}$.  We now switch the conditional gates back on, in the first of the stages announced above, and optimize them over local Gaussian unitaries so as to eliminate this outcome dependence.  We begin with displacements alone, $\hat{U}_{G_i}(\mathcal{M})=\hat{D}(\alpha_i (\mathcal{M}))$, as shown in Fig.~\ref{optimal_displacement_circuit}. In this scheme, displacements with parameters $\alpha_i (\mathcal{M})\quad (i=1,2)$ determined by $\mathcal{M}$ are applied to the two post-measurement modes after the homodyne detection. The gate parameters generally also depend on the protocol parameters $\lambda$ and $\theta$, but we display only the dependence on $\mathcal{M}$ to emphasize the measurement-based feed-forward structure.

As a result, the success probability remains the same as in Eq.~\eqref{eq:basic_success}, while the fidelity becomes independent of $\mathcal{M}$ and is given by the optimal value \begin{equation}\label{eq:opt_fid}
    F_{\mathrm{opt}}(\lambda,\theta) = \frac{\sqrt{\left(1-\lambda ^2\right) \left(1-\lambda ^2 \cos^2\theta \right)}}{1-\tfrac{3}{8}\lambda^{2}-\tfrac{1}{2}\lambda^{2}\cos\theta-\tfrac{1}{8}\lambda^{2}\cos 2\theta}. 
\end{equation}
The optimal fidelity $F_{\mathrm{opt}}(\lambda,\theta)$ as functions of $\theta$ for several values of $\lambda$ are shown together with $F_{\mathrm{avg}}(\lambda,\theta)$ in Fig.~\ref{optimal_displacement_fidelity}.

The monotonic growth of $P$ with $\theta$ noted above admits a simple physical interpretation: as $\theta$ increases, more of the input state is reflected into the measured mode, making the outcomes more sensitive to the sign of the squeezing. In contrast, stronger reflection reduces the contribution of the initial state to the remaining mode, leading to a decrease in the fidelity of the post-measurement state. 

Since $P$ and $F$ vary oppositely with $\theta$, the score function of Eq.~\eqref{eq:score} admits a non-trivial maximum. We therefore write
\begin{equation}\label{score_general}
  \textrm{Score function }=  \overline{P\cdot F} \leq \mathcal{B},
\end{equation}
and aim to determine the bounds $\mathcal{B}$ for different restrictions on Gaussian operations, resources, and measurements.

For a finite $d$-dimensional system, the bound obtained in Ref.~\cite{Lim:2025aa} is ${\cal B} = d^2 \xi^4/2$, where $\xi$ denotes the largest Schmidt coefficient. However, this result cannot be directly applied to our setting, since $d$ diverges for TMSVs with fixed $\xi$, yielding only a trivial bound.

\subsection{Local squeezing improves fidelity}
\label{sub:local_squeezing}

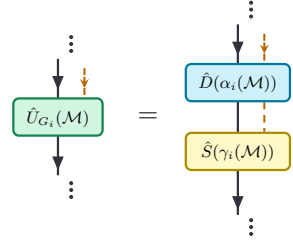
\begin{figure}
 \begin{center}
 \scalebox{0.7}{
\begin{tikzpicture}[x=1cm,y=1cm]

\tikzset{ccline/.style={draw=qoFB,line width=1.1pt,
  dash pattern=on 4pt off 3pt,dash phase=0.21pt}}
\begin{scope}[yshift=-0.65cm]
  \draw[W](0,1.1)--(0,-1.1);
  \draw[ccline](0.5,0.923)--(0.5,0.76);
  \draw[draw=qoFB,line width=1.1pt,->,>=stealth](0.5,0.76)--(0.5,0.5491);
  \draw[ccline](0.5,0.5491)--(0.5,0.3726);
  \node[draw=qoUG,fill=qoUGL,line width=1.2pt,rounded corners=3pt,
        minimum width=48pt,minimum height=20pt,font=\small,inner sep=3pt]
        at(0,0){$\hat{U}_{G_i}(\mathcal{M})$};
  \ArrowD{0}{0.76}
  \ArrowD{0}{-0.76}
  \foreach \y in {1.25,1.38,1.51}{\filldraw[qoWire](0.25,\y)circle(0.8pt);}
  \foreach \y in {-1.25,-1.38,-1.51}{\filldraw[qoWire](0.25,\y)circle(0.8pt);}
\end{scope}
\node[font=\Large] at(1.7,-0.65){$=$};
\begin{scope}[xshift=3.4cm]
  \draw[W](0,1.1)--(0,-2.5);
  %% Orange: entry then D->S inter-gate
  \draw[ccline](0.5,0.923)--(0.5,0.76);
  \draw[draw=qoFB,line width=1.1pt,->,>=stealth](0.5,0.76)--(0.5,0.5491);
  \draw[ccline](0.5,0.5491)--(0.5,0.3726);
  \draw[ccline](0.5,-0.3726)--(0.5,-0.9274);
  %% Gates after orange so fill covers orange inside boxes
  \node[draw=qoDisp,fill=qoDispL,line width=1.2pt,rounded corners=3pt,
        minimum width=56pt,minimum height=20pt,font=\small,inner sep=3pt]
        at(0,0){$\hat{D}(\alpha_i(\mathcal{M}))$};
  \node[draw=qoSq,fill=qoSqL,line width=1.2pt,rounded corners=3pt,
        minimum width=56pt,minimum height=20pt,font=\small,inner sep=3pt]
        at(0,-1.3){$\hat{S}(\gamma_i(\mathcal{M}))$};
  \ArrowD{0}{0.76}
  \ArrowD{0}{-2.16}
  \foreach \y in {1.25,1.38,1.51}{\filldraw[qoWire](0.25,\y)circle(0.8pt);}
  \foreach \y in {-2.60,-2.73,-2.86}{\filldraw[qoWire](0.25,\y)circle(0.8pt);}
\end{scope}
\end{tikzpicture}}
\end{center}
\caption{\label{squeezing_circuit}Circuit with additional local squeezing on both modes, built on top of the optimal displacement scheme (Fig.~\ref{optimal_displacement_circuit}). The local Gaussian gates are parametrized by the set of measurement outcomes $\mathcal{M}=\{x_3,x_4\}$. Since the measurement-based displacements $\hat{D}\left(\alpha_i(\mathcal{M})\right)\quad(i=1,2)$ render the two subsystems symmetric, both modes are squeezed by the same outcome-independent factor $\gamma_1=\gamma_2=\gamma_\mathrm{opt}$, fixed by the input squeezing $\lambda$ and the beamsplitter angle $\theta$. This corresponds to choosing $l=0$ and $\hat{U}_{G_i}(\mathcal{M})=\hat{D}(\alpha_i (\mathcal{M}))\hat{S}(\gamma_i (\mathcal{M}))$ in the general circuit.}
\end{figure}

Having fixed the measurement-based displacement, we next activate the local squeezing operations of the general circuit, with parameters $\gamma_i \;(i=1,2)$ applied to each post-measurement mode, as shown in Fig.~\ref{squeezing_circuit}.
Unlike the conditional displacements, the optimal squeezing parameters depend only on the protocol parameters $\lambda$ and $\theta$ and are independent of the measurement outcomes $\mathcal{M}$.

The fidelity, defined in the same way as in the previous subsection, takes the form
\begin{equation}
    F_\mathrm{squ}(\lambda,\theta,\gamma)=\frac{(1-\gamma^{2})\sqrt{\left(1-\lambda ^{2}\right) \left(1-\lambda ^{2} \cos^{2}\theta\right)}}{1-\left(\lambda  \cos ^2\frac{\theta }{2}-\gamma  \lambda  \sin ^2\frac{\theta}{2}\right)^2}.
    \label{eq:squ_fid}
\end{equation}
Here $\theta$ is the beamsplitter angle and $\gamma$ quantifies the strength of the symmetric local squeezing applied to both modes. 
We first optimize the protocol with respect to the local squeezing, which enters only through the fidelity and leaves the success probability unchanged. Consequently, it can be fixed independently prior to optimizing the success probability-fidelity score function $\overline{P\cdot F}$, which we now refer to as the score function. Differentiating Eq.~\eqref{eq:squ_fid} with respect to $\gamma$ and setting the derivative to zero yields, for each fixed pair $(\lambda,\theta)$, the optimal squeezing parameter 
\begin{equation}
    \gamma_\mathrm{opt}(\lambda,\theta)=-\frac{\left(\sqrt{1-\lambda ^2 \cos ^2\theta }-\sqrt{1-\lambda ^2}\right)^{2}}{\lambda ^2 \left(1-\cos ^2\theta\right)}.
    \label{eq:opt_squeezing}
\end{equation}

Inserting $\gamma = \gamma_{\mathrm{opt}}(\lambda,\theta)$ into Eq.~\eqref{eq:squ_fid} maximizes the fidelity over the local squeezing. The resulting expression depends on $\lambda$ and $\theta$ only through the success probability $P$ of Eq.~\eqref{eq:basic_success}, and collapses to the compact form 
\begin{equation}
    F_{\mathrm{squ}}(\lambda,\theta,\gamma_{\mathrm{opt}}) =\frac{2 \sin(P\pi)}{1+\sin(P\pi)} =: F_{\mathrm{squ}}(P),
\end{equation}
as we derive in Appendix~\ref{app:succ}. At fixed $\lambda$, the monotonicity of $P$ in $\theta$ makes the map $\theta \mapsto P$ invertible, with accessible range $\left[1/2, P(\lambda,\pi)\right]$, so that $P$ may be used in place of $\theta$ as the independent variable. This states the tradeoff in its sharpest form: once the local squeezing has been optimized, the fidelity is a strictly decreasing function of the success probability alone, independently of how a given $P$ was realized. 

The score function in Eq.~\eqref{eq:score} therefore becomes 
\begin{equation}
    \overline{P\cdot F_\mathrm{squ}}=P\times F_{\mathrm{squ}}(P)=\frac{2 P\sin(P\pi)}{1+\sin(P\pi)},
    \label{eq:opt_squ_sco_p}
\end{equation}
so that the joint optimization over $\gamma$ and $\theta$ collapses to a one-dimensional problem in $P$. Equation~\eqref{eq:opt_squ_sco_p} has a unique stationary point $P_\mathrm{opt}$, fixed by a transcendental condition derived in Appendix~\ref{app:succ}. The optimization is analytic throughout; only the root of that condition is evaluated numerically, giving $P_\mathrm{opt}=0.713804$. 

\begin{figure} \label{glocc_score}
\center
\includegraphics[scale=0.9]{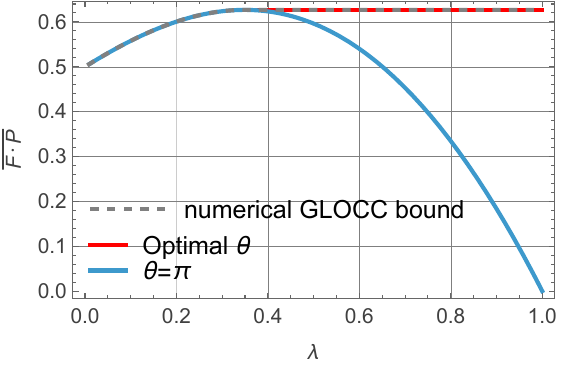}
\caption{Score  $\overline{P\cdot F}$ as a function of the input squeezing $\lambda$ for different local Gaussian strategies. The blue curve corresponds to the simple strategy where the beamsplitter angle is fixed to $\theta = \pi$ and no additional squeezing is applied, so that the initial state is fully reflected and locally measured, and the fidelity is evaluated with respect to two vacuum modes. The red solid line shows the constant optimal score $\overline{P\cdot F_{\mathrm{squ}}} = 0.626829$ achieved for $\lambda \geq \lambda_c= 0.349066$ by the locally squeezed scheme of Fig.~\ref{squeezing_circuit}. The gray curve represents the numerically obtained score for a fully general local Gaussian circuit, indicating that the optimized locally squeezed scheme already saturates the corresponding $\textrm{GLOCC}_1$ bound.}\end{figure}

At this optimal point, the score function attains its maximum value $\overline{P\cdot F_\mathrm{squ}}=P_\mathrm{opt}\times F_{\mathrm{squ}}(P_\mathrm{opt})=\frac{2 P_\mathrm{opt} \sin(P_\mathrm{opt}\pi)}{1+\sin(P_\mathrm{opt}\pi)}=0.626829$, reached at the beamsplitter angle $\theta_\mathrm{opt}$ fixed implicitly by 
\begin{equation}
    \frac{\lambda(1-\cos \theta_\mathrm{opt})}{1-\lambda^2 \cos \theta_\mathrm{opt}}=-\cos(\pi P_\mathrm{opt})=0.622306.
\end{equation}
Since the accessible values of $P$ are bounded above by $P(\lambda,\pi)$, this optimum is reachable only beyond the threshold $\lambda_c$ defined by $P(\lambda_c,\pi)=P_\mathrm{opt}$, namely $\lambda_c=\tan(\pi P_\mathrm{opt})-\sec(\pi P_\mathrm{opt})=0.349066$. For $\lambda_c\leq \lambda < 1$ the optimized score is therefore independent of $\lambda$, as indicated by the red solid line in Fig.~\ref{glocc_score}.

For weakly squeezed initial states with $\lambda < \lambda_c$, the stationary point lies outside the attainable range. As the score $\overline{P \cdot F_\mathrm{squ}}$ increases monotonically below $P_\mathrm{opt}$ and $P$ increases with $\theta$, the maximum is attained at the boundary $\theta=\pi$ rather than at an interior stationary point, giving the closed-form score 
\begin{equation}
    \overline{P \cdot F_\mathrm{squ}}=(1-\lambda^2)\left[\frac{1}{2} +\frac{1}{\pi} \arcsin \left(\frac{2\lambda}{1+\lambda^2}\right)\right].
\end{equation}
This clearly demonstrates that local squeezing provides a significant advantage in the regime of larger input squeezing, whereas for weakly entangled inputs, the optimal strategy essentially coincides with the displacement-only circuit.

\subsection{Numerical verification}\label{glocc}

\begin{figure}
 \begin{center}
 \scalebox{0.7}{
\begin{tikzpicture}[x=1cm,y=1cm]

\tikzset{ccline/.style={draw=qoFB,line width=1.1pt,
  dash pattern=on 4pt off 3pt,dash phase=0.21pt}}

\begin{scope}[yshift=-1.85cm]
  \draw[W](0, 1.1)--(0,-1.1);
  \draw[ccline] (0.5,0.923)--(0.5,0.76);
  \draw[draw=qoFB,line width=1.1pt,->,>=stealth] (0.5,0.76)--(0.5,0.5491);
  \draw[ccline] (0.5,0.5491)--(0.5,0.3726);
  \node[draw=qoUG,fill=qoUGL,line width=1.2pt,rounded corners=3pt,
        minimum width=48pt,minimum height=20pt,font=\small,inner sep=3pt]
        at(0, 0.0){$\hat{U}_{G_i}(\mathcal{M})$};
  \ArrowD{0}{0.76}
  \ArrowD{0}{-0.76}
  \foreach \y in {1.25, 1.38, 1.51}{
    \filldraw[qoWire](0.25,\y)circle(0.8pt);}
  \foreach \y in {-1.25, -1.38, -1.51}{
    \filldraw[qoWire](0.25,\y)circle(0.8pt);}
\end{scope}

\node[font=\Large] at(1.7,-1.85){$=$};

\begin{scope}[xshift=3.4cm]
  \draw[W](0, 1.1)--(0,-4.8);
  \draw[ccline] (0.5,0.923)--(0.5,0.76);
  \draw[draw=qoFB,line width=1.1pt,->,>=stealth] (0.5,0.76)--(0.5,0.5491);
  \draw[ccline] (0.5,0.5491)--(0.5,0.3726);
  \draw[ccline] (0.5,-0.3726)--(0.5,-0.9274);
  \draw[ccline] (0.5,-1.6726)--(0.5,-2.2274);
  \draw[ccline] (0.5,-2.9726)--(0.5,-3.3165);
  \draw[draw=qoFB,line width=1.1pt,->,>=stealth] (0.5,-3.3165)--(0.5,-3.5274);
  \draw[ccline] (0.5,-3.5274)--(0.5,-4.2726);
  \node[draw=qoDisp,fill=qoDispL,line width=1.2pt,rounded corners=3pt,
        minimum width=56pt,minimum height=20pt,font=\small,inner sep=3pt]
        at(0, 0.0){$\hat{D}(\alpha_i(\mathcal{M}))$};
  \node[draw=qoRot,fill=qoRotL,line width=1.2pt,rounded corners=3pt,
        minimum width=56pt,minimum height=20pt,font=\small,inner sep=3pt]
        at(0,-1.3){$\hat{R}(\varphi_i(\mathcal{M}))$};
  \node[draw=qoSq,fill=qoSqL,line width=1.2pt,rounded corners=3pt,
        minimum width=56pt,minimum height=20pt,font=\small,inner sep=3pt]
        at(0,-2.6){$\hat{S}(\gamma_i(\mathcal{M}))$};
  \node[draw=qoRot,fill=qoRotL,line width=1.2pt,rounded corners=3pt,
        minimum width=56pt,minimum height=20pt,font=\small,inner sep=3pt]
        at(0,-3.9){$\hat{R}(\phi_i(\mathcal{M}))$};
  \ArrowD{0}{0.76}
  \ArrowD{0}{-4.46}
  \foreach \y in {1.25, 1.38, 1.51}{
    \filldraw[qoWire](0.25,\y)circle(0.8pt);}
  \foreach \y in {-4.95, -5.08, -5.21}{
    \filldraw[qoWire](0.25,\y)circle(0.8pt);}
\end{scope}

\end{tikzpicture}}
\end{center}
\caption{
\label{general_gaussian_unitary}
General local Gaussian circuit employed for the non-destructive discrimination of two-mode squeezed vacua. 
 Each mode is subjected to a single-mode Gaussian operation decomposed into local squeezing $\gamma_i(\mathcal{M})$, displacement $\alpha_i(\mathcal{M})$, and phase rotations $\phi_i(\mathcal{M}),\; \varphi_i(\mathcal{M})$. 
 The displacement is fixed by the measurement-based prescription that centers the post-measurement state, while the remaining parameters are numerically optimized.
 This circuit provides the general local Gaussian processing stage considered in Sec.~\ref{glocc}.}
\end{figure}
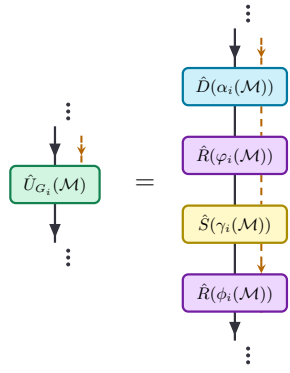

We investigate how far one can improve the non-destructive discrimination of two-mode squeezed vacua within the framework of local Gaussian operations, and numerically obtain an upper bound on the score function in this setting. Returning to the full local Gaussian circuit of Fig.~\ref{general_gaussian_circuit}, we now optimize over all of its parameters. Each conditional single-mode operation $\hat{U}_{G_i}(\mathcal{M})$ is decomposed into a displacement $\alpha_i(\mathcal{M})$, a squeezing $\gamma_i(\mathcal{M})$, and two phase rotations $\varphi_i(\mathcal{M}),\, \phi_i(\mathcal{M})$, as shown in Fig.~\ref{general_gaussian_unitary}. As with our previous examples, we fix the local displacement to the measurement-based prescription that shifts the post-measurement state to the origin and numerically optimize the remaining Gaussian parameters, thereby bounding the performance attainable under local Gaussian processing.

The numerically obtained bound, displayed in Fig.~\ref{glocc_score}, coincides within our numerical precision over the entire range of input squeezing we consider with the score achieved by the locally squeezed protocol of Fig.~\ref{squeezing_circuit}. This suggests that this relatively simple scheme already achieves the optimal performance within our class of local Gaussian operations with one adaptive stage, and we therefore conjecture that it realizes the $\textrm{GLOCC}_1$ bound for this configuration.

\section{Non-destructive discrimination with pre-shared entanglement}\label{sec:preshared}
Pre-shared entanglement can serve as a useful resource for improving the non-destructive discrimination of two-mode squeezed vacua. In this section, we investigate how the entanglement shared between Alice and Bob prior to the protocol can enhance the achievable performance. The behaviour of the protocol depends qualitatively on the relation between the amount of pre-shared entanglement and the squeezing of the input state. Depending on this relation, two distinct regimes emerge. In some parameter ranges, the discrimination task becomes essentially trivial once sufficient pre-shared entanglement is available. In other regimes, however, the problem remains non-trivial and requires a more careful analysis. These two cases are discussed separately in the following subsections.

\subsection{Trivial cases}
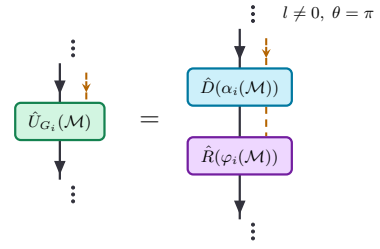
\begin{figure}
 \begin{center}
 \scalebox{0.7}{
\begin{tikzpicture}[x=1cm,y=1cm]
\tikzset{ccline/.style={draw=qoFB,line width=1.1pt,
  dash pattern=on 4pt off 3pt,dash phase=0.21pt}}
\node[anchor=west,font=\small\itshape,text=black]
  at(4.1,1.38){$l \neq 0,\; \theta = \pi$};
\begin{scope}[yshift=-0.65cm]
  \draw[W](0,1.1)--(0,-1.1);
  \draw[ccline](0.5,0.923)--(0.5,0.76);
  \draw[draw=qoFB,line width=1.1pt,->,>=stealth](0.5,0.76)--(0.5,0.5491);
  \draw[ccline](0.5,0.5491)--(0.5,0.3726);
  \node[draw=qoUG,fill=qoUGL,line width=1.2pt,rounded corners=3pt,
        minimum width=48pt,minimum height=20pt,font=\small,inner sep=3pt]
        at(0,0){$\hat{U}_{G_i}(\mathcal{M})$};
  \ArrowD{0}{0.76}
  \ArrowD{0}{-0.76}
  \foreach \y in {1.25,1.38,1.51}{\filldraw[qoWire](0.25,\y)circle(0.8pt);}
  \foreach \y in {-1.25,-1.38,-1.51}{\filldraw[qoWire](0.25,\y)circle(0.8pt);}
\end{scope}
\node[font=\Large] at(1.7,-0.65){$=$};
\begin{scope}[xshift=3.4cm]
  \draw[W](0,1.1)--(0,-2.5);
  \draw[ccline](0.5,0.923)--(0.5,0.76);
  \draw[draw=qoFB,line width=1.1pt,->,>=stealth](0.5,0.76)--(0.5,0.5491);
  \draw[ccline](0.5,0.5491)--(0.5,0.3726);
  \draw[ccline](0.5,-0.3726)--(0.5,-0.9274);
  \node[draw=qoDisp,fill=qoDispL,line width=1.2pt,rounded corners=3pt,
        minimum width=56pt,minimum height=20pt,font=\small,inner sep=3pt]
        at(0,0){$\hat{D}(\alpha_i(\mathcal{M}))$};
  \node[draw=qoRot,fill=qoRotL,line width=1.2pt,rounded corners=3pt,
        minimum width=56pt,minimum height=20pt,font=\small,inner sep=3pt]
        at(0,-1.3){$\hat{R}(\varphi_i(\mathcal{M}))$};
  \ArrowD{0}{0.76}
  \ArrowD{0}{-2.16}
  \foreach \y in {1.25,1.38,1.51}{\filldraw[qoWire](0.25,\y)circle(0.8pt);}
  \foreach \y in {-2.60,-2.73,-2.86}{\filldraw[qoWire](0.25,\y)circle(0.8pt);}
\end{scope}
\end{tikzpicture}}
\end{center}
\caption{\label{trivial_circuit}Trivial non-destructive discrimination circuit with fully reflective beamsplitters: the entire initial state is measured, and a measurement-based local rotation ($\varphi_i(\mathcal{M})$), equal to $\frac{\pi}{2}$ for the outcome ($-\lambda$) to flip the sign of the pre-shared entanglement and match it to the measured initial state, and $0$ otherwise.}
\end{figure}

We first consider the circuit in which all beamsplitters are set to full reflection, $\theta=\pi$, so that the entire input is routed to the measurement stage and the output modes are inherited from the pre-shared state without any mixing, as illustrated in Fig.~\ref{trivial_circuit}. We call this the trivial circuit, in the sense that it performs measurement and re-preparation rather than genuine local Gaussian processing; whether it is in fact optimal is a separate question, settled in Sec.~\ref{non_trivial_regime}.
The pre-shared entanglement can then be routed entirely towards preserving the output fidelity in the remaining modes. 

\begin{figure}[t] \label{score_trivial_circuit}
\center
\includegraphics[scale=0.17]{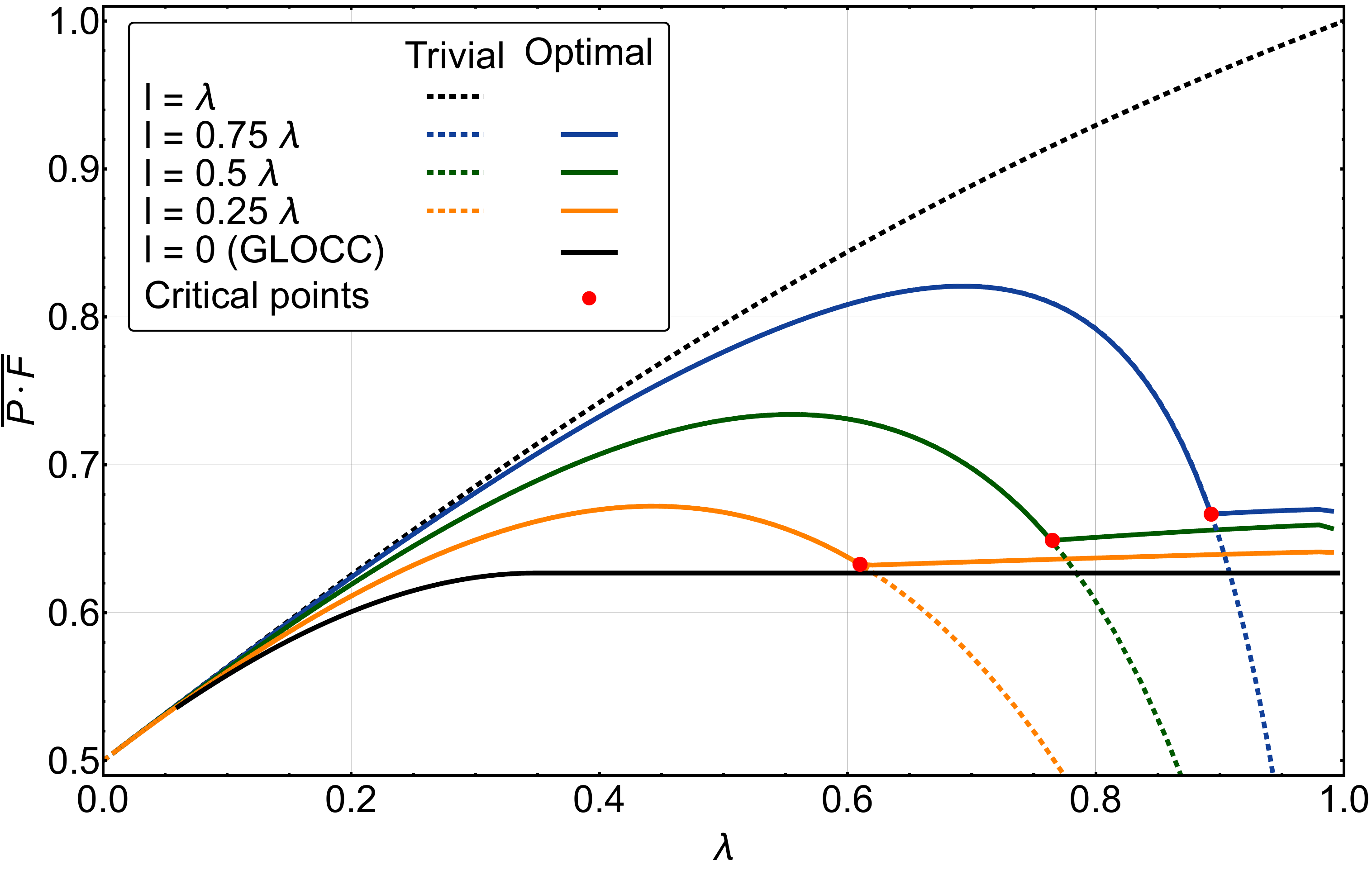}
\caption{Score of the pre-shared entanglement-assisted protocol as a function of the input squeezing $\lambda$, for several ratios $l/\lambda$. The solid black line indicates the $\textrm{GLOCC}_1$ bound obtained without pre-shared entanglement ($l=0$). Dashed curves show the trivial non-destructive discrimination circuit of Fig.~\ref{trivial_circuit}, in which the entire input state is measured and re-prepared from the pre-shared entanglement, with the matched case $l=\lambda$ drawn as a black dashed line. For $l<\lambda$, the trivial score eventually decreases to zero as $\lambda \to 1$, which lies outside the displayed score range.  Solid coloured curves show the numerically optimized general local Gaussian circuit of Fig.~\ref{pre_shared_general_gaussian_circuit}: the trivial curve remains optimal up to a critical value of $\lambda$, marked by the scatter points for $l/\lambda=0.25, 0.5, 0.75$, beyond which the optimal strategy enters the non-trivial regime.}
\end{figure}

Following the homodyne measurement, we apply a measurement-based local rotation $\varphi(\mathcal{M})$, defined by $\varphi(\mathcal{M})=\frac{\pi}{2}$ when the measurement outcome $\mathcal{M}$ is assigned to $\ket{-\lambda}$ by the decision rule, and $\varphi(\mathcal{M})=0$ otherwise.
The pre-shared entanglement is always prepared with a positive squeezed parameter $+l$, whereas the initial state takes the values $\pm\lambda$ with equal probability. The conditional rotation $\varphi(\mathcal{M})$ therefore flips the sign of the pre-shared entanglement precisely for the $-\lambda$ outcome, matching it to the measured initial state and completing the discrimination.
For this trivial configuration, the beamsplitters are fully reflective and the conditional rotation $\varphi(\mathcal{M})$ is fixed by the rule above, so the protocol is characterized entirely by the 
input squeezing $\lambda$ and the pre-shared entanglement $l$. 
Since the entire input is measured, the success probability follows from Eq.~\eqref{eq:basic_success} at $\theta=\pi$ and is independent of $l$.
The output state, is the pre-shared state with its sign matched to the measured input by the conditional rotation, so the fidelity is the same for both signs and reduces to the overlap $\left|\langle l | \lambda \rangle \right|^2 =  \frac{(1-l^2)(1-\lambda^2)}{(1-l\lambda)^2}$.
The score is therefore, 
\begin{equation}
    \overline{P\cdot F}\left(l,\lambda\right)=\left(\frac{\tan ^{-1}\left(\frac{2 \lambda }{1-\lambda ^2}\right)}{\pi }+\frac{1}{2}\right)\frac{(1-l^2)(1-\lambda^2)}{(1-l\lambda)^2},
\end{equation}
which reduces to the success probability alone for a matched pre-shared state, $l=\lambda$.
The behaviour of this score is shown by the dashed curves in Fig.~\ref{score_trivial_circuit} for several ratios $l/\lambda$. 
For a matched pre-shared state the score increases monotonically with $\lambda$ and approaches unity in the infinitely squeezed limit, where the two signs of the input become perfectly distinguishable by homodyne measurement. 
This score lies above the $\textrm{GLOCC}_1$ bound of Sec.~\ref{glocc}, drawn as the solid black line in Fig.~\ref{score_trivial_circuit}, over the whole range of $\lambda$, showing that a matched pre-shared state surpasses the limit of local Gaussian strategies without pre-shared entanglement even though no mixing is performed at the beamsplitters.
For a mismatched state with $l<\lambda$, the score still exceeds the $\textrm{GLOCC}_1$ bound at small $\lambda$, where the overlap between the pre-shared and input states remains large.
As $\lambda$ grows at fixed $l/\lambda$, however, the overlap decreases and the growth of the success probability cannot compensate for the loss. The score therefore falls back below the $\textrm{GLOCC}_1$ bound and vanishes as $\lambda\to1$, where the numerator of the overlap carries $(1-\lambda^2)$ while the denominator stays finite.

With a matched pre-shared state, the trivial circuit therefore exceeds the $\textrm{GLOCC}_1$ bound across the whole range of input squeezing, even though it performs no mixing at the beamsplitters. Preparing such a state, however, requires a large amount of entanglement whenever the input is strongly squeezed, which may be experimentally demanding, and the mismatched case $l<\lambda$ is therefore the one of practical relevance. At large $\lambda$, however, the trivial circuit falls below the $\textrm{GLOCC}_1$ bound --- worse than using no entanglement at all. To benefit from the pre-shared entanglement here, one must therefore mix it with the input at the beamsplitters.

\subsection{Non-trivial regime}\label{non_trivial_regime}

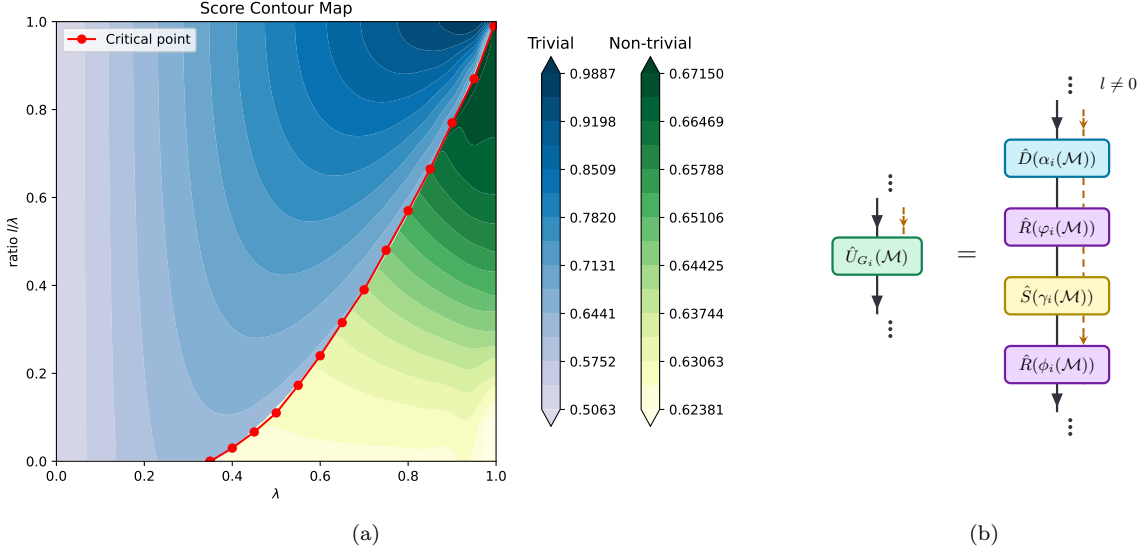
\begin{figure*}[t]
    \subfloat[]{
    \includegraphics[scale=0.5]{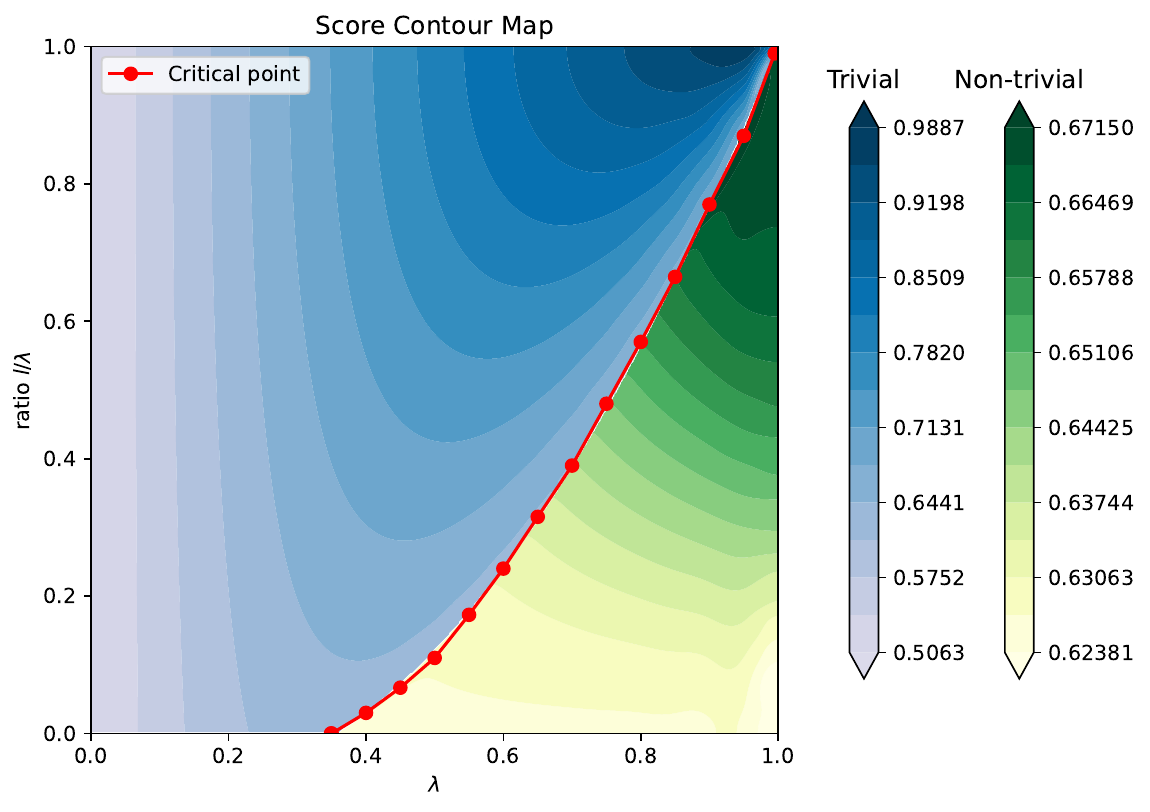}% Here is how to import EPS art
    \label{density_plot}
   }
    \hspace{1cm}\subfloat[]{\raisebox{1cm}{
 \scalebox{0.7}{
\begin{tikzpicture}[x=1cm,y=1cm]
\tikzset{ccline/.style={draw=qoFB,line width=1.1pt,
  dash pattern=on 4pt off 3pt,dash phase=0.21pt}}

%% Label beside dots on RHS (top dots at y=1.51, no yshift on RHS)
\node[anchor=west,font=\small\itshape,text=black]
  at(4.1,1.38){$l \neq 0$};

%% LHS shifted to match 4-gate RHS centre (y=-1.85)
\begin{scope}[yshift=-1.85cm]
  \draw[W](0,1.1)--(0,-1.1);
  \draw[ccline](0.5,0.923)--(0.5,0.76);
  \draw[draw=qoFB,line width=1.1pt,->,>=stealth](0.5,0.76)--(0.5,0.5491);
  \draw[ccline](0.5,0.5491)--(0.5,0.3726);
  \node[draw=qoUG,fill=qoUGL,line width=1.2pt,rounded corners=3pt,
        minimum width=48pt,minimum height=20pt,font=\small,inner sep=3pt]
        at(0,0){$\hat{U}_{G_i}(\mathcal{M})$};
  \ArrowD{0}{0.76}
  \ArrowD{0}{-0.76}
  \foreach \y in {1.25,1.38,1.51}{\filldraw[qoWire](0.25,\y)circle(0.8pt);}
  \foreach \y in {-1.25,-1.38,-1.51}{\filldraw[qoWire](0.25,\y)circle(0.8pt);}
\end{scope}

\node[font=\Large] at(1.7,-1.85){$=$};

%% RHS: full 4-gate stack
\begin{scope}[xshift=3.4cm]
  \draw[W](0,1.1)--(0,-4.8);
  %% Orange entry
  \draw[ccline](0.5,0.923)--(0.5,0.76);
  \draw[draw=qoFB,line width=1.1pt,->,>=stealth](0.5,0.76)--(0.5,0.5491);
  \draw[ccline](0.5,0.5491)--(0.5,0.3726);
  %% Inter-gate segments
  \draw[ccline](0.5,-0.3726)--(0.5,-0.9274);
  \draw[ccline](0.5,-1.6726)--(0.5,-2.2274);
  \draw[ccline](0.5,-2.9726)--(0.5,-3.3165);
  \draw[draw=qoFB,line width=1.1pt,->,>=stealth](0.5,-3.3165)--(0.5,-3.5274);
  \draw[ccline](0.5,-3.5274)--(0.5,-4.2726);
  %% Gates (after orange so fill covers)
  \node[draw=qoDisp,fill=qoDispL,line width=1.2pt,rounded corners=3pt,
        minimum width=56pt,minimum height=20pt,font=\small,inner sep=3pt]
        at(0,0){$\hat{D}(\alpha_i(\mathcal{M}))$};
  \node[draw=qoRot,fill=qoRotL,line width=1.2pt,rounded corners=3pt,
        minimum width=56pt,minimum height=20pt,font=\small,inner sep=3pt]
        at(0,-1.3){$\hat{R}(\varphi_i(\mathcal{M}))$};
  \node[draw=qoSq,fill=qoSqL,line width=1.2pt,rounded corners=3pt,
        minimum width=56pt,minimum height=20pt,font=\small,inner sep=3pt]
        at(0,-2.6){$\hat{S}(\gamma_i(\mathcal{M}))$};
  \node[draw=qoRot,fill=qoRotL,line width=1.2pt,rounded corners=3pt,
        minimum width=56pt,minimum height=20pt,font=\small,inner sep=3pt]
        at(0,-3.9){$\hat{R}(\phi_i(\mathcal{M}))$};
  \ArrowD{0}{0.76}
  \ArrowD{0}{-4.46}
  \foreach \y in {1.25,1.38,1.51}{\filldraw[qoWire](0.25,\y)circle(0.8pt);}
  \foreach \y in {-4.95,-5.08,-5.21}{\filldraw[qoWire](0.25,\y)circle(0.8pt);}
\end{scope}
\end{tikzpicture}}}
    \label{pre_shared_general_gaussian_circuit}}
\label{results}
\caption{(a) The landscape of solutions with pre-shared entanglement with two distinct solutions: measurement and repreparation (trivial) and optimised mixing at each beamsplitter (non-trivial). Moving up the $y-$axis on the graph always increases the value of the score function -- pre-shared entanglement is always a useful resource in GLOCC discrimination. (b) General local Gaussian circuit for the non-destructive discrimination of two-mode squeezed vacua assisted by pre-shared entanglement $l$. The single-mode Gaussian gates retain the same measurement-based parametrization as in Fig.~\ref{general_gaussian_unitary}, with the sole modification that the ancillary modes $3,4$ are prepared in the entangled state $|l\rangle_{34}$ rather than the vacuum.
}
\end{figure*}

We now turn to the regime in which the input squeezing is sufficiently large that the fully reflective circuit is no longer optimal and a more general local Gaussian circuit is required. To characterize the achievable performance, we numerically optimize the general pre-shared entanglement-assisted local Gaussian circuit of Fig.~\ref{pre_shared_general_gaussian_circuit}, treating the beamsplitter angle $\theta$ and the conditional local Gaussian gates as free parameters. The optimized score is shown as the solid coloured curves of Fig.~\ref{score_trivial_circuit}, each obtained at a fixed ratio $l/\lambda$. Inspecting the resulting optimal parameters, we find that for a mismatched pre-shared state, $l<\lambda$, the optimum is attained at $\theta=\pi$ up to a critical input squeezing, marked by the scatter points on each curve. Below this value the optimized curves coincide with the trivial score of the preceding subsection, so that measuring the entire input and re-preparing from the pre-shared state is optimal in this range even though the two states do not match.

Beyond this critical value the optimal beamsplitter angle departs from $\theta=\pi$ and the conditional local squeezing acquires a non-zero value. The input state is then partially transmitted rather than fully measured, and the output fidelity is supported by mixing the residual input with the pre-shared entanglement rather than by the overlap $\left|\langle l | \lambda \rangle \right|^2$ alone. The optimized score consequently separates from the trivial one, which decays towards zero as $\lambda\to1$, and instead stays above the $\textrm{GLOCC}_1$ bound of Sec.~\ref{glocc} throughout. We thus find that the optimized protocol exceeds the $\textrm{GLOCC}_1$ limit over the entire range of $\lambda$ for any non-zero pre-shared entanglement. 

The same transition is displayed over the full parameter space in Fig.~\ref{density_plot}. Whereas Fig.~\ref{score_trivial_circuit} shows cuts at fixed $l/\lambda$, the density plot resolves the optimized score on the $(\lambda, l/\lambda)$ plane, on which the critical points of the former trace out the boundary separating the two regimes. The boundary moves to larger $\lambda$ as $l/\lambda$ increases, and closes off at $l=\lambda$, for which the fully reflective circuit is optimal over the whole range of input squeezing. Across both regimes the optimized score increases monotonically with $l$ at fixed $\lambda$. For small input squeezing the score is uniformly small and this dependence is not resolved on the scale of Fig.~\ref{density_plot}; a zoomed-in comparison at $\lambda=0.1$ and $0.2$ is given in  Appendix~\ref{app:small_squeezing_regime}, where the fully reflective circuit is optimal throughout.

Once the non-trivial circuits are included, pre-shared entanglement therefore improves the attainable score at every value of the input squeezing, and does so above the level accessible without it.

\section{Conclusion}\label{sec:conc}
In this work, we extended the notion of non-destructive state discrimination from the discrete-variable setting to the continuous-variable setting. This extension required a ground-up reformulation because the original bounds given in terms of the Hilbert space dimension are unsuitable for continuous-variable systems. We focused on the local Gaussian regime where we discriminated TMSVs with $\textrm{GLOCC}_1$ operations. We obtained analytical results for the success probability-fidelity tradeoff without pre-shared entangled resources, and we obtained numerical results when one has access to pre-shared entanglement. In the latter case, we identified two distinct regions: first, a trivial measurement-repreparation circuit which involves an optimal local Gaussian measurement of the input signal states and comparing the fidelity of the pre-shared entangled state with the initial state (i.e., no mixing at the beamsplitters); second, a non-trivial regime where we obtain non-trivial optimal parameters for mixing at the beamsplitters.

We considered all possible local Gaussian unitaries to optimise the success probability-fidelity score function. In particular, we noticed that local squeezing can improve the fidelity between a two-mode ground state and a TMSV. While the utility of local squeezing has been studied in the context of improving teleportation fidelity \cite{Bowen:2003aa,PhysRevA.78.062340}, our results extend this to the standard fidelity figure of merit. We analysed the setting where one has limited access to pre-shared entanglement, $l$, and found the optimal circuit for any given value of initial squeezing $\lambda$ in terms of the fraction $\frac{l}{\lambda}$. As expected, we found that pre-shared entanglement can always improve the score function, and this monotonically increases with increasing $l$.

Some open questions and interesting avenues to explore remain. An analytical derivation of the fully adaptive GLOCC bound remains lacking. It is known that in certain quantum information tasks there exists a strict gap between general-dyne measurements and adaptive general-dyne measurements \cite{chen2026sampleoptimallearningbosonicgaussian}. We conjecture that additional ancillary vacuum states and adaptive stages will allow for modest improvements in the success probability-fidelity tradeoff, but we leave this question for future work. Moreover, it is known that non-Gaussian resources are useful in state discrimination of certain Gaussian states \cite{PhysRevA.78.022320,Moran2026nearoptimalcoherent}, although for the discrimination of general Gaussian states, it is still unknown what the optimal Gaussian measurement is. Understanding this would allow us to extend our work to a broader class of Gaussian quantum states. Although not all non-Gaussian operations are currently practical, some are performed routinely, such as photon counting, and others, like the cubic phase gate and selective number arbitrary phase (SNAP) gate, have seen significant progress in recent years \cite{PhysRevLett.115.137002,Park_2024}. They are still of theoretical relevance for understanding and realising the fundamental limits of such tasks.

\acknowledgements{This work was supported by the Institute of Information \& Communications Technology Planning \& Evaluation (IITP) grant (No. 2022-0-00463, No. IITP-2024-RS-2024-00437284, and No. IITP-2025-RS-2025-02283189), funded by the Korea government (MSIT). J.M. and H.K. are supported by KIAS individual grant numbers QP088702 (J.M.) CG085302 (H.K.) at the Korea Institute for Advanced Study. M.J.S. and M.-S.C. were supported by National Research Foundation of Korea (Grant Nos. RS-2023-NR068116 and 2023R1A2C1005588) and the Ministry of Education through the BK21 Program.}

\appendix

\section{Comparison of homodyne and heterodyne}
\label{app: comparison_of_homodyne_and_heterodyne}
\begin{figure}
    \centering

    %(a)
    \subfloat[\label{fig:hete_succ_90}] {%\includegraphics[width=0.45\textwidth]
    \includegraphics[width=0.235\textwidth]
    {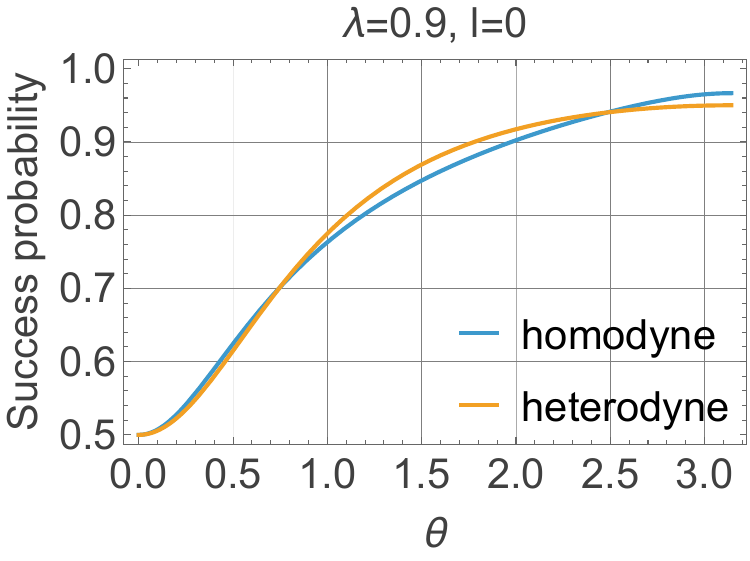}
    }
    \hfill
    \subfloat[\label{fig:hete_succ_95}]{%\includegraphics[width=0.45\textwidth]
    \includegraphics[width=0.235\textwidth]
    {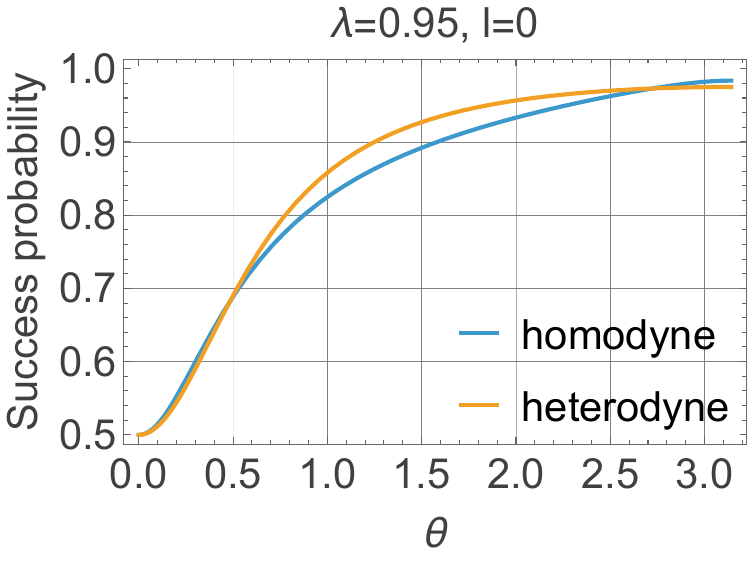}}
    
    \vspace{0.3cm}    %(c)
    
    \subfloat[\label{fig:hete_score_90}]{%\includegraphics[width=0.45\textwidth]
    \includegraphics[width=0.235\textwidth]{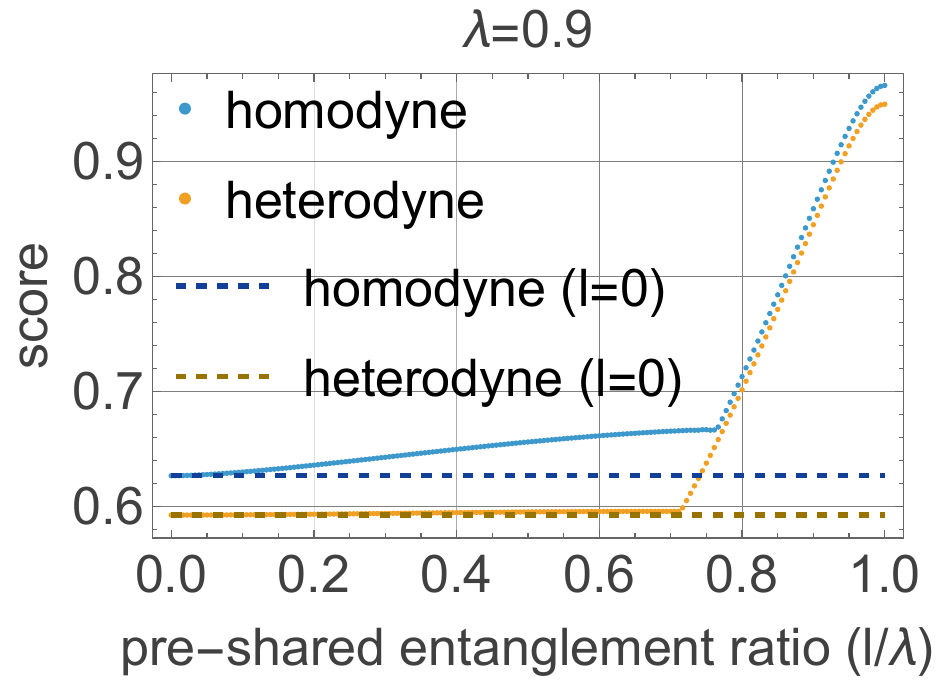}}
    \hfill
    \subfloat[\label{fig:hete_score_95}]{%\includegraphics[width=0.45\textwidth]
    \includegraphics[width=0.235\textwidth]
    {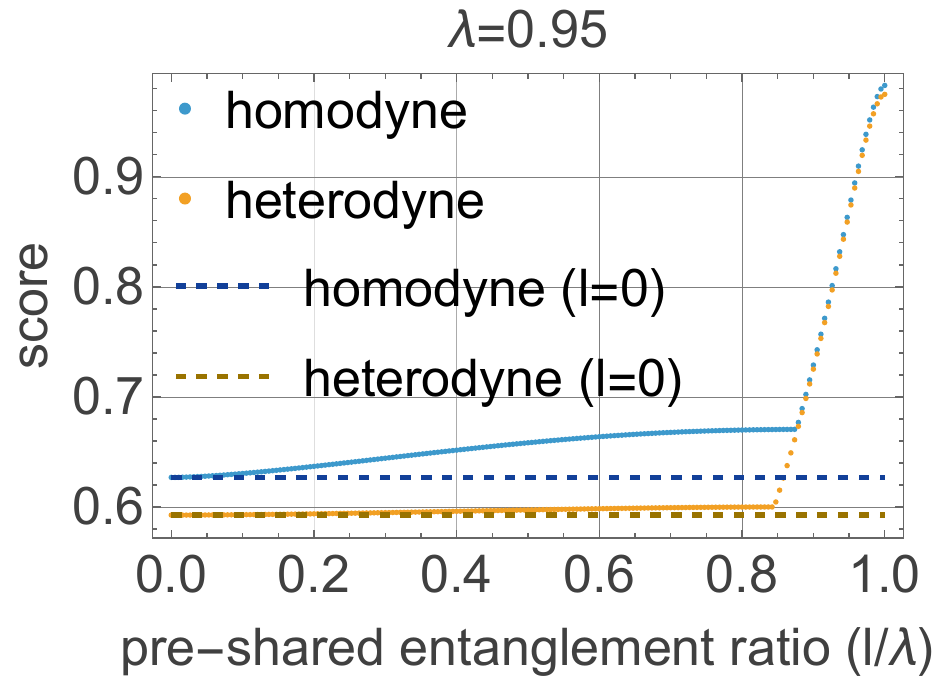}} 
    \caption{\label{fig:hete_hom}Homodyne detection is not always optimal for minimum-error discrimination of two Gaussian states with Gaussian measurements, but it remains optimal for the score function (Eq.~\eqref{eq:score}) in this work. At $\lambda=0.9$ in (a) we see that there are regions of the beamsplitter angle where heterodyne outperforms homodyne, but in its corresponding graph of the score function in (c) we see that homodyne remains optimal. A similar effect for $\lambda=0.95$ is shown in (b) and (d).}
\end{figure}

Here we demonstrate numerically that while homodyne detection is not always optimal (among local Gaussian measurements) for minimum-error discrimination of two TMSVs, it is still optimal for the fidelity-success probability tradeoff introduced in the main text. In Figs.~\ref{fig:hete_succ_90} and ~\ref{fig:hete_succ_95} we find regions of $\lambda$ and the beamsplitter transmissivity $\theta$ where heterodyne detection achieves better success probability than homodyne detection. In the corresponding graphs in Figs.~\ref{fig:hete_score_90} and ~\ref{fig:hete_score_95}, respectively, we see that the score function $\overline{P \cdot F}$ is still saturated by homodyne detection.

\section{Success probability as the variable of the locally squeezed protocol}
\label{app:succ}
\subsection{Monotonicity and range of the success probability}
Throughout the main text, we use the success probability of Eq.~\eqref{eq:basic_success} in place of the beamsplitter angle. Here we establish the properties that justify this substitution. 
We first verify that the argument of the arcsine,
\begin{equation}
    u(\lambda,\theta)=\frac{\lambda(1-\cos\theta)}{1-\lambda^2 \cos \theta},
\end{equation}
remains within the domain of the arcsine on the physical range $0<\lambda<1,\;0\leq \theta \leq \pi$. The denominator obeys $1-\lambda^2 \cos \theta \geq 1-\lambda^2 > 0$, so $u$ is non-negative and regular, while
\begin{equation}
    1-u = \frac{(1-\lambda)(1+\lambda \cos \theta)}{1-\lambda^2 \cos \theta}>0.
\end{equation}
Thus $u \in [0,1)$, where the arcsine is differentiable with positive derivative, and the monotonicity of $P$ is therefore inherited from that of $u$, whose partial derivatives are 
\begin{equation}
    \frac{\partial u}{\partial \lambda}=\frac{(1-\cos\theta)(1+\lambda^2 \cos\theta)}{(1-\lambda^2 \cos \theta)^2},\;\; 
    \frac{\partial u}{\partial \theta}=\frac{\lambda(1-\lambda^2)\sin\theta}{(1-\lambda^2 \cos \theta)^2}.
\end{equation}
Both are manifestly non-negative on this domain, so $P$ increases monotonically with each of $\lambda$ and $\theta$. The range of $P$ is then fixed by its endpoint values. 
Evaluating the argument at the boundaries gives $u=0$ for $\theta=0$ and for $\lambda=0$, and $1$ for $\lambda=1$, so that 
\begin{equation}
    P|_{\theta=0} = P|_{\lambda=0} = \frac{1}{2},\quad P|_{\lambda=1} = 1.    
\end{equation}
The first two correspond to a fully transmissive beamsplitter and to a product input, for which the two candidate states induce identical outcome distributions and no better strategy than random guessing exists.
At fixed $\lambda$, therefore, $\theta \mapsto P$ is a continuous strictly increasing bijection onto $\left[\frac{1}{2}, P(\lambda, \pi)\right]$ with, 
\begin{equation}
    P(\lambda,\pi)=\frac{1}{2}+\frac{1}{\pi}\arcsin \left(\frac{2\lambda}{1+\lambda^2}\right),
    \label{eq:p_max}
\end{equation}
which is the statement used in Sec.~\ref{sub:local_squeezing}.

\subsection{Reduction of the squeezing-optimized fidelity}
The change of variables established above eliminates $\theta$ in favour of $P$ at fixed $\lambda$. We now show that the fidelity of the locally squeezed scheme of Sec.~\ref{sub:local_squeezing}, where both modes are squeezed by a common factor $\gamma$. When evaluated at the optimal value $\gamma_{\mathrm{opt}}$ of Eq.~\eqref{eq:opt_squeezing}, the fidelity of this scheme loses its separate dependence on $\lambda$. Consequently, it becomes a function of $P$ alone. Since the local squeezing affects only the fidelity and leaves the success probability unchanged, $\gamma$ can be fixed independently of the remaining parameters; the stationarity condition $\frac{\partial F_\mathrm{squ}}{\partial \gamma}=0$ applied to Eq.~\eqref{eq:squ_fid} yields the $\gamma_\mathrm{opt}(\lambda, \theta)$ quoted in Eq.~\eqref{eq:opt_squeezing}. Carrying out this substitution in Eq.~\eqref{eq:squ_fid} and simplifying gives the maximized fidelity in the explicit form
\begin{equation}\begin{split}F_{\mathrm{squ}}(\lambda,\theta,\gamma_{\mathrm{opt}})=&\frac{2\sqrt{(1-\lambda^2)(1-\lambda^2\cos^2\theta)}}{\lambda^2(1-\cos\theta)^2}\Bigl[1-\lambda^2\cos\theta \\&-\sqrt{(1-\lambda^2)(1-\lambda^2\cos^2\theta)}\Bigr].\end{split}\label{eq:opt_squ_fid}\end{equation}
Every $\lambda$ and $\theta$ dependence here is carried by the numerator $\lambda(1-\cos\theta)$ and the denominator $1-\lambda^2 \cos \theta$ of $u$. Indeed, expanding both sides gives the algebraic identity
\begin{equation}\begin{split}
    (1-\lambda^2 \cos \theta)^2 - \lambda^2 (1-\cos \theta)^2 &\\= (1-\lambda^2)(1-\lambda^2 \cos^2 \theta).
\end{split}\end{equation}
which states precisely that
\begin{align}
    \frac{\sqrt{(1-\lambda^2)(1-\lambda^2\cos^2\theta)}}{\lambda(1-\cos\theta)}
    &=\sqrt{\frac{1}{u^2}-1}, \nonumber \\
    \frac{1-\lambda^2\cos\theta-\sqrt{(1-\lambda^2)(1-\lambda^2\cos^2\theta)}}{\lambda(1-\cos\theta)}
    & \nonumber \\
    =\frac{1}{u}&-\sqrt{\frac{1}{u^2}-1}.
    \label{eq:u_factors}
\end{align}
Multiplying the two relations of Eq.~\eqref{eq:u_factors} and restoring the overall factor of two reproduces Eq.~\eqref{eq:opt_squ_fid}, so that
\begin{equation}
    F_\mathrm{squ}(u)=\frac{2\sqrt{1-u^2}\left(1-\sqrt{1-u^2}\right)}{u^2},
    \label{eq:fid_u}
\end{equation}
a function of $u$ alone, with no residual dependence on $\lambda$ or $\theta$ separately.
Since $u$ is the argument of the arcsine in Eq.~\eqref{eq:basic_success}, it equals $-\cos(\pi P)$, and $\sqrt{1-u^2}=\sin(\pi P)$ with the positive sign fixed by $P\in\left[\frac{1}{2},1\right)$. Eq.~\eqref{eq:fid_u} therefore becomes
\begin{equation}
    F_\mathrm{squ}(P)=\frac{2\sin(\pi P)\left[1-\sin(\pi P)\right]}{1-\sin^2(\pi P)}=\frac{2\sin(\pi P)}{1+\sin(\pi P)}.
    \label{eq:opt_squ_fid_p}
\end{equation}
This is the reduced fidelity quoted in Sec.~\ref{sub:local_squeezing}, and the reduction is exact on the whole domain $0<\lambda<1$, $0<\theta\leq\pi$. As a consistency check, consider the fully reflective case $\theta=\pi$, for which $\gamma_\mathrm{opt}\to 0$ by Eq.~\eqref{eq:opt_squeezing}. Substituting $P(\lambda,\pi)$ of Eq.~\eqref{eq:p_max} into Eq.~\eqref{eq:opt_squ_fid_p}, with $\sin[\pi P(\lambda,\pi)]=(1-\lambda^2)/(1+\lambda^2)$, gives $F_\mathrm{squ}(\lambda,\pi,\gamma_\mathrm{opt})=1-\lambda^2$, the fidelity of a two-mode vacuum with the input TMSV, as it must since the post-measurement state is the vacuum in this limit. Note that Eqs.~\eqref{eq:squ_fid} and \eqref{eq:opt_squeezing} admit no such simplification, since the residual $\lambda$ dependence cancels only once $\gamma_\mathrm{opt}$ has been inserted; they are therefore quoted in terms of $(\lambda,\theta,\gamma)$ throughout.

\subsection{Optimal success probability and the input-squeezing threshold}

With Eq.~\eqref{eq:opt_squ_sco_p} the score function of the locally squeezed scheme is a function of $P$ alone, so that
\begin{equation}
    \frac{\mathrm{d}}{\mathrm{d}P}\overline{P\cdot F_\mathrm{squ}}
    =\frac{2\left[\sin^2(\pi P)+\sin(\pi P)+\pi P\cos(\pi P)\right]}{\left[1+\sin(\pi P)\right]^2}.
    \label{eq:score_derivative}
\end{equation}
Stationarity therefore requires
\begin{equation}
    h(P)\equiv\sin^2(\pi P)+\sin(\pi P)+\pi P\cos(\pi P)=0,
    \label{eq:transcendental}
\end{equation}
in which $P$ appears both inside and outside trigonometric functions. Equation~\eqref{eq:transcendental} is transcendental and has no solution in closed form. Its endpoint values are $h(1/2)=2$ and $h(1)=-\pi$, so a root exists in $\left(\frac{1}{2},1\right)$ by continuity; a numerical scan confirms that $h$ changes sign only once there, so the root is unique and is the global maximum of the score function on this interval. We obtain $P_\mathrm{opt}=0.713804$, at which $\overline{P\cdot F_\mathrm{squ}}=0.626829$ and $-\cos(\pi P_\mathrm{opt})=0.622306$.

The threshold $\lambda_c$ quoted in the main text follows from $P(\lambda_c,\pi)=P_\mathrm{opt}$. Using Eq.~\eqref{eq:p_max}, this reads $2\lambda_c/(1+\lambda_c^2)=-\cos(\pi P_\mathrm{opt})$, a quadratic whose root in $(0,1)$ is
\begin{equation}
    \lambda_c=\frac{1-\sin(\pi P_\mathrm{opt})}{-\cos(\pi P_\mathrm{opt})}
    =\tan(\pi P_\mathrm{opt})-\sec(\pi P_\mathrm{opt})=0.349066.
    \label{eq:lambda_c}
\end{equation}
For $\lambda<\lambda_c$ the accessible interval $\left[\frac{1}{2},P(\lambda,\pi)\right]$ lies entirely below $P_\mathrm{opt}$, where Eq.~\eqref{eq:score_derivative} is positive, so the score is maximized at the endpoint $\theta=\pi$.

\section{Small squeezing regime}
\label{app:small_squeezing_regime}

\begin{figure}
\center
\includegraphics[scale=0.5]{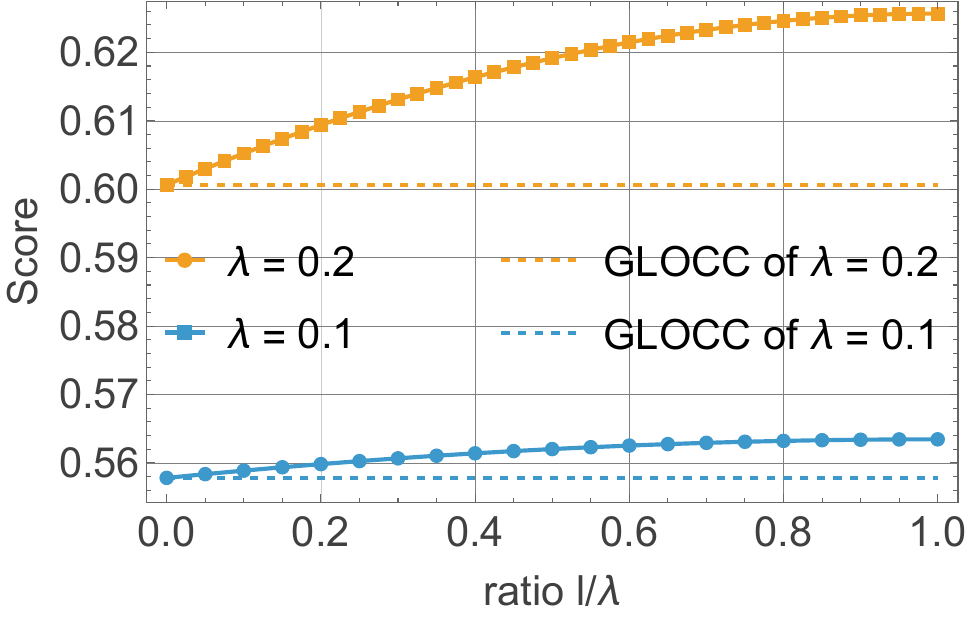}
\caption{Zoomed-in view of the achievable score in the small input-squeezing regime ($\lambda=0.1, 0.2$) as a function of the pre-shared entanglement $l$. The data points are obtained numerically using the trivial fully reflective circuit of Fig.~\ref{trivial_circuit}, which remains optimal in this regime. The dashed lines indicate the corresponding GLOCC bounds (without pre-shared entanglement), showing that any non-zero pre-shared entanglement already provides an advantage. Although the overall score is low for small $\lambda$, this improvement becomes clearly visible in the zoomed-in plot.}
\label{small_lambda}
\end{figure}

In the small input-squeezing regime of the pre-shared entanglement-assisted protocol discussed in Sec .~\ref {non_trivial_regime}, the overall score is relatively small, making it difficult to resolve the dependence on the pre-shared entanglement in the density plot. The density plot in Fig.~\ref{density_plot} summarizes the overall behavior of the protocol across the full parameter range, including the transition between trivial and non-trivial regimes. However, in the small-squeezing region, the variation with respect to the pre-shared entanglement is less visually pronounced. To make this behavior more transparent, we present a zoomed-in analysis for representative values $\lambda=0.1$ and $0.2$ in Fig.~\ref{small_lambda}. In this small-squeezing regime, the trivial fully reflective circuit of Fig.~\ref{trivial_circuit} already achieves optimal performance and is therefore used throughout.

As shown in Fig.~\ref{small_lambda}, the achievable score increases monotonically with the pre-shared entanglement $l$ for both values of $\lambda$. Moreover, the score always exceeds the corresponding GLOCC bound, indicated by dashed lines, which is independent of $l$ due to the absence of pre-shared entanglement. The gap from this bound becomes larger as the pre-shared entanglement approaches the input-state squeezing. Although the absolute improvement is modest in this regime due to the small input squeezing, this trend is clearly resolved in the zoomed-in representation, thereby complementing the results in the main text.

\bibliographystyle{apsrev4-2-titles}
\bibliography{NonDestBib.bib}

\end{document}